\documentclass[superscriptaddress, aps, prl, reprint, twocolumn, amsmath, amssymb, showpacs]{revtex4-1}

\usepackage{graphicx}  
\usepackage{mathptmx}
\usepackage{epstopdf}
\usepackage{dcolumn}   
\usepackage{bm}        
\usepackage{amssymb}   
\usepackage{amsmath}   
\usepackage{amsthm}
\usepackage{chemarrow}
\usepackage{color}
\usepackage{xcolor} 
\usepackage{mathrsfs}
\usepackage{float}
\usepackage{physics}
\usepackage{bbold}
\usepackage{bbm}
\usepackage[toc,page]{appendix}
\usepackage{longtable}
\usepackage{multirow}
\usepackage{utfsym}
\usepackage{upgreek}
\usepackage{epstopdf}
\usepackage{ulem}
\usepackage{graphicx}
\usepackage{lipsum}

\usepackage[a4paper, colorlinks=true, linkcolor=blue, citecolor=blue,
pdfauthor={ },
pdftitle={ },
pdfsubject={ },
pdfkeywords={ }]{hyperref}
\usepackage[version=4]{mhchem}
\usepackage{ulem,fancyvrb}

\theoremstyle{plain}
\newtheorem*{theorem*}{Theorem}

\begin{document}

\title{Dark Matter Search with a Resonantly-Coupled Hybrid Spin System}

    \author{Kai Wei}
    \email[]{These authors contributed equally to this work}
    \affiliation{
	School of Instrumentation Science and Opto-electronics Engineering, Beihang University, Beijing, 100191, China}
    \affiliation{
	Hangzhou Extremely Weak Magnetic Field Major Science and Technology Infrastructure Research Institute, Hangzhou, 310051, China}
    \affiliation{
    Hefei National Laboratory, Hefei 230088, China}
	
	\author{Zitong Xu}
	\email[]{These authors contributed equally to this work}
    \affiliation{
		School of Physical and Mathematical Sciences, Nanyang Technological University, Singapore, 639798, Singapore}
	\affiliation{
		School of Instrumentation Science and Opto-electronics Engineering, Beihang University, Beijing, 100191, China}
	\affiliation{
		Hangzhou Extremely Weak Magnetic Field Major Science and Technology Infrastructure Research Institute,
Hangzhou, 310051, China}
	
	\author{Yuxuan He}
	\affiliation{School of Physics and State Key Laboratory of Nuclear Physics and Technology, Peking University, Beijing 100871, China}
	\author{Xiaolin Ma}
	\affiliation{School of Physics and State Key Laboratory of Nuclear Physics and Technology, Peking University, Beijing 100871, China}
	
	\author{Xing Heng}
	\affiliation{
		School of Instrumentation Science and Opto-electronics Engineering, Beihang University, Beijing, 100191, China}
	\affiliation{
		Hangzhou Extremely Weak Magnetic Field Major Science and Technology Infrastructure Research Institute,
Hangzhou, 310051, China}
	
	\author{Xiaofei Huang}
	\affiliation{
		School of Instrumentation Science and Opto-electronics Engineering, Beihang University, Beijing, 100191, China}
	\affiliation{
		Hangzhou Extremely Weak Magnetic Field Major Science and Technology Infrastructure Research Institute,
Hangzhou, 310051, China}

	\author{Wei Quan}
	\affiliation{
		School of Instrumentation Science and Opto-electronics Engineering, Beihang University, Beijing, 100191, China}
	\affiliation{
		Hangzhou Extremely Weak Magnetic Field Major Science and Technology Infrastructure Research Institute,
Hangzhou, 310051, China}
	
	\author{Wei Ji}
	\email[Corresponding author: ]{wei.ji@pku.edu.cn}
    \affiliation{School of Physics and State Key Laboratory of Nuclear Physics and Technology, Peking University, Beijing 100871, China}
	\affiliation{Johannes Gutenberg University, Mainz 55128, Germany}
	\affiliation{Helmholtz-Institute, GSI Helmholtzzentrum fur Schwerionenforschung, Mainz 55128, Germany}
	
	\author{Jia Liu}
	\email[Corresponding author: ]{jialiu@pku.edu.cn}
	\affiliation{School of Physics and State Key Laboratory of Nuclear Physics and Technology, Peking University, Beijing 100871, China}
	\affiliation{Center for High Energy Physics, Peking University, Beijing 100871, China}
	
	\author{Xiao-Ping Wang}
	\affiliation{School of Physics, Beihang University, Beijing 100191, China}
    \affiliation{Beijing Key Laboratory of Advanced Nuclear Materials and Physics, Beihang University, Beijing 100191, China}

	\author{Dmitry Budker}
	\affiliation{Johannes Gutenberg University, Mainz 55128, Germany}
	\affiliation{Helmholtz-Institute, GSI Helmholtzzentrum fur Schwerionenforschung, Mainz 55128, Germany}
	\affiliation{Department of Physics, University of California at Berkeley, Berkeley, California 94720-7300, USA}

       \author{Jiancheng Fang}
	\email[Corresponding author: ]{fangjiancheng@buaa.edu.cn}
	\affiliation{
		School of Instrumentation Science and Opto-electronics Engineering, Beihang University, Beijing, 100191, China}
	\affiliation{
		Hangzhou Extremely Weak Magnetic Field Major Science and Technology Infrastructure Research Institute,
Hangzhou, 310051, China}
    \affiliation{
    Hefei National Laboratory, Hefei 230088, China}
	
\begin{abstract}
Recent advances in tabletop quantum sensor technology have enabled searches for nongravitational interactions of dark matter (DM). Traditional axion DM experiments rely on sharp resonance, resulting in extensive scanning time to cover a wide mass range. 
In this work, we present a broadband approach in an alkali-${}^{21}$Ne spin system. 
We identify two distinct hybrid spin-coupled regimes: a self-compensation (SC) regime at low frequencies and a hybrid spin resonance (HSR) regime at higher frequencies.
By utilizing these two distinct regimes, we significantly enhance the bandwidth of ${}^{21}$Ne nuclear spin compared to conventional nuclear magnetic resonance, while maintaining competitive sensitivity. We present a comprehensive broadband search for axion-like dark matter, covering 5 orders of magnitude of Compton frequencies range within $[10^{-2}, \, 10^3]$\,Hz. We set new constraints on the axion dark matter interactions with neutrons and protons, accounting for the effects of DM stochasticity. For the axion-neutron coupling, our results reach a low value of $|g_{ann}|\le 3\times 10^{-10}$ in the frequency range $[2\times 10^{-2}, \, 4]$\,Hz surpassing astrophysical limits and providing the strongest laboratory constraints in the $[10, \, 100]$\,Hz range. 
For the axion-proton coupling, we offer the best terrestrial constraints for the frequency ranges $[2\times 10^{-2}, \, 5]$\,Hz and $[16, \, 7\times 10^{2}]$\,Hz.
\end{abstract}

\maketitle

\noindent \textbf{\large {Introduction}} 

\noindent There is evidence of dark matter (DM) from numerous astrophysical and cosmological observations. However, its nature is yet to be understood, which is one of the most important challenges in modern physics.
Ultralight pseudoscalar particles, such as axions or axionlike particles (ALPs)~\cite{osti_1306469, particle2020review}, are well-motivated dark matter candidates, which can be produced via the ``misalignment mechanism'' to obtain the correct relic abundance~\cite{duffy2009axions, marsh2016axion,preskill1983cosmology,abbott1983cosmological,dine1983not}. The axion is particularly well motivated as it also solves the so-called strong-CP problem by introducing the Peccei-Quinn global symmetry~\cite{peccei1977cp,  weinberg1978new, wilczek1978problem, vafa1984parity}. It (weakly) couples to other particles~\cite{golub1994neutron} due to the spontaneous breaking at high energy, making it difficult to detect in experiments. 
In this article, we refer to both axions and ALPs as ``axions'', so that the coupling to photons, gluons and fermions can span a wide range of parameter space which has been explored in various astrophysical and laboratory experiments~\cite{raffelt1990astrophysical, graham2015experimental, safronova2018search}.

Regarding the coupling to nucleons, the axion gradient field can be measured as a pseudomagnetic field coupling to the nuclei. In the non-relativistic limit, the gradient interaction can be described by the Hamiltonian,
	\begin{align}
		\mathcal{H} =  g_{\rm aNN} \bm{\nabla} a \cdot \bf{I} ,
		\label{eq.Hami-nucleus}
	\end{align}
where $\bf{I}$ is the spin of the nucleus which includes fraction of spins of the neutron and proton,and $g_{\rm aNN}$ is the effective axion coupling to the nucleus. 
Therefore, the pseudomagnetic field due to the axion DM is $\bm{b}_a \equiv g_{\rm aNN} \bm{\nabla} a /\gamma_{\rm n}$, where $\gamma_{\rm n}$ is the gyromagnetic ratio of the nuclear spin.

Recent experiments are using various nuclei, including the liquid state experiments using proton and $^{13}$C~\cite{wu2019search} and gas state sensors based on alkali-noble pairs using $^3$He~\cite{bloch2023constraints, lee2022laboratory, gavilan2024searching} and $^{129}$Xe~\cite{jiang2021search}. These terrestrial experiments explore the nuclear precession frequency range from \,\rm{$\mu$}Hz to hundreds of kHz, corresponding to the axion mass range from 10$^{-22}$ to 10$^{-10}$~eV. In the frequency range from approximately 0.01\,Hz to 10\,Hz, the $^3$He comagnetometer~\cite{lee2022laboratory}, together with our previous work through the ChangE collaboration (Coupled Hot Atom eNsembles for liGht dark mattEr) on the bandwidth-enhanced nuclear magnetic resonance (NMR) magnetometer~\cite{xu2024constraining} sets limits on the axion-neutron coupling to be lower than 10$^{-9}\, \rm{GeV}^{-1}$. A recent $^3$He comagnetometer~\cite{gavilan2024searching} provides slightly weaker limits in the 0.01\,Hz to 10\,Hz range but can extend these limits down to $\mathcal{O}(10^{-7})$ Hz frequencies.

Typically, to expand the search range for DM with gas-state quantum sensors, a bias magnetic field is applied. The resonance frequency is scanned by adjusting the external magnetic field to match the Larmor frequency with the axion Compton frequency. 
In \cite{bloch2023constraints}, a spin-exchange-relaxation-free (SERF) magnetometer used with a leading field that is sufficiently high to break the SERF regime, leading to increased spin-exchange relaxation. In \cite{jiang2021search}, a spin-based amplifier is applied, where the resonance frequency is scanned by adjusting the external magnetic field to match the Larmor frequency with the axion Compton frequency. 
Resonant searches benefit from high Q factors at the resonant frequency but suffer from narrow bandwidth, typically at the level of 0.01\,Hz for Xe~\cite{jiang2021search} and Ne~\cite{xu2024constraining}. Thus, to search for an axion with unknown mass, one needs to scan the frequency with steps corresponding to the narrow bandwidth to cover the target frequency range, rendering resonant searches to be time-consuming.

In this work, we apply the hybrid spin manipulation method, first proposed and demonstrated in the K-${}^3$He system~\cite{kornack2002dynamics}, to conduct a broadband search using the K-Rb-${}^{21}$Ne system. In ${}^{21}$Ne system, we identify two distinct regimes: one resembling the self-compensation (SC) regime observed in the ${}^3$He system, and another at a higher frequency which we term as the hybrid spin resonance (HSR) mode. The SC and HSR regimes for the K-${}^3$He system are nearly degenerate, limiting the bandwidth coverage in the dark matter search. These regimes exhibit a significantly broader bandwidth compared to typical NMR magnetometers, while maintaining competitive sensitivity in both cases. They allow for the transformation of the conventional narrow-band resonant search into a broadband search for axion dark matter, benefiting from extended experiment durations. We utilize the K-Rb-${}^{21}$Ne sensor to search for axion dark matter spanning 5 orders of magnitude, [0.01, 1000] Hz in the frequency range. After analyzing the data and rigorously examining potential candidates, we establish new constraints on the axion-neutron and axion-proton couplings.
\\

\noindent \textbf{\large{RESULTS}}

\noindent \textbf{{Experimental Setup}}

We use a K-Rb-$^{21}$Ne hybrid thermal spin ensembles in which the minority potassium atoms are used for ``hybrid optical pumping" \cite{babcock2003hybrid} and the mixed spin ensemble of alkali metal and noble gas is operated in a regime where the two spin species are resonantly-coupled and under the influence of the magnetic field from each other. The alkali metal gas and noble gas undergo hybrid spin resonance under an appropriate external magnetic field.

\begin{figure}[htb]
    \centering
\includegraphics[width= 0.95  \linewidth]{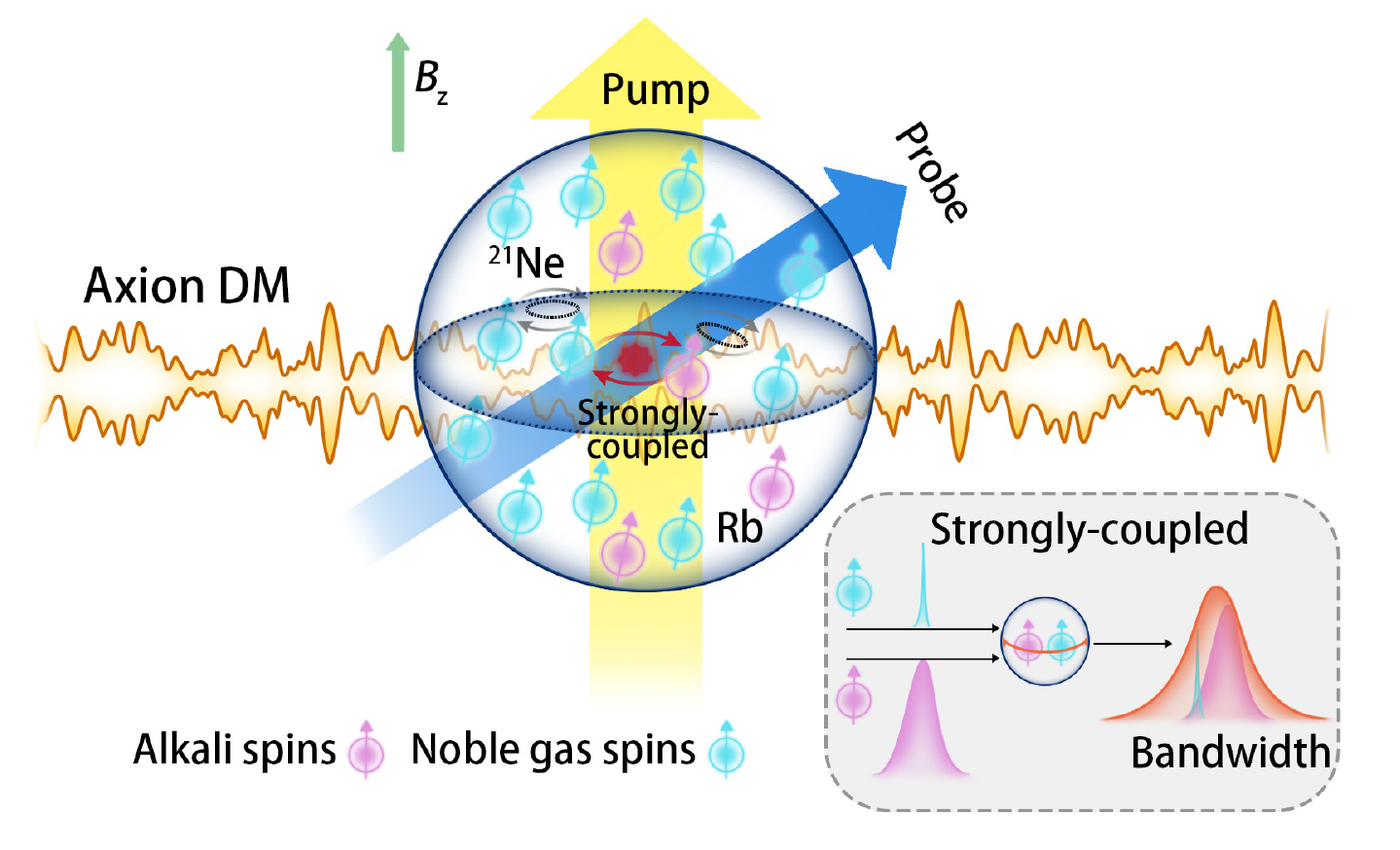}
\caption{The principle of HSR and dark matter detection: (1) Pump light polarizes the alkali electron spins, while spin-exchange collisions with alkali spins polarize the noble-gas nuclear spins. (2) The axion dark matter produced a pseudomagnetic field acting on nuclear spins. The precession of nuclear spins is transferred to alkali spins via Fermi-contact interactions. (3) The resulting precession of the alkali spins is read out using  optical rotation of the probe light. Normally, the measurement bandwidth of noble-gas nuclear spins is narrow due to their small relaxation rate $\Gamma_{\rm n}$ (blue peak), which is several orders of magnitude smaller than that of alkali spins (purple peak). However, in the HSR regime, the noble-gas spins become resonantly-coupled with alkali spins. As a result, the interactions with alkali spins significantly broaden the bandwidth of the noble-gas spins (red peak).}
\label{Fig.prin} 
\end{figure}

The experimental setup comprises a spherical cell with a 11.4\,mm internal diameter that holds a small droplet of K and Rb metal, along with $70\%$ isotope-enriched $^{21}$Ne. To prevent the spontaneous emission transition of alkali atoms in the excited state and the generation of resonant photons with random spin polarization, we used N$_2$ gas as a quenching gas. The alkali atoms are spin-polarized with circularly-polarized light tuned to the potassium D1 line. The hybrid optical pumping is used to improve the hyperpolarization efficiency for $^{21}$Ne nuclear spins and reduce the polarization gradient of alkali spins, which is achieved by optically pumping the optically-thin potassium vapor to avoid strong resonant absorption of the pump light\,\cite{wei2022constraints, Wei:2022ggs}. The spin angular momentum is transferred from the ``minority'' K atoms to the Rb atoms and the $^{21}$Ne nuclear spins via spin-exchange collisions. Additionally, $^{21}$Ne gas also acts as a buffer gas to reduce the rapid wall-collision relaxation of alkali spins.  The precession of polarized $^{21}$Ne nuclear spins is monitored with optically-thick Rb spins, read out by detecting optical rotation of linearly-polarized probe light detuned to low-frequency side of  the D1 line of Rb. The above processes can be summarized as an illustration picture in Fig.~\ref{Fig.prin}.
\\

\noindent \textbf{Hybrid Spin Dynamics}

The spin interactions between hyperpolarized $^{21}$Ne atoms and spin-polarized alkali atoms are dominated by Fermi-contact interactions (FCI), which  can be characterized as effective fields ${\bf{{B}^{e/n}}}=\lambda M^{\rm{e/n}}_0 {\bf{P^{e/n}}}$ due to the magnetization of the other spin species, where $\lambda$ is the Fermi-contact enhancement factor, $M^{\rm{e/n}}_0$ is magnetization of alkali (noble-gas) atoms for fully polarization, ${\bf{P^{e/n}}}$ is the collective spin polarization of alkali (noble-gas) spin ensembles. Therefore, the total fields experienced by alkali spins and noble-gas spins are  $\textbf{B}^\textbf{e}_{\rm tot}={\bf{{B}^{n}}} + {\bf{B}}$ and  $\textbf{B}^\textbf{n}_{\rm tot}={\bf{{B}^{e}}} + {\bf{B}}$ respectively, where ${\bf{B}}$ is the applied magnetic field. Typically, the gyromagnetic ratio of alkali spins is much larger than that of noble-gas nuclear spins. Consequently, the Larmor frequency of alkali spins at a given field is usually orders of magnitude higher than that of noble-gas spins. 

\begin{figure*}[htb]
    \centering
    \includegraphics[width= 0.7 \textwidth]{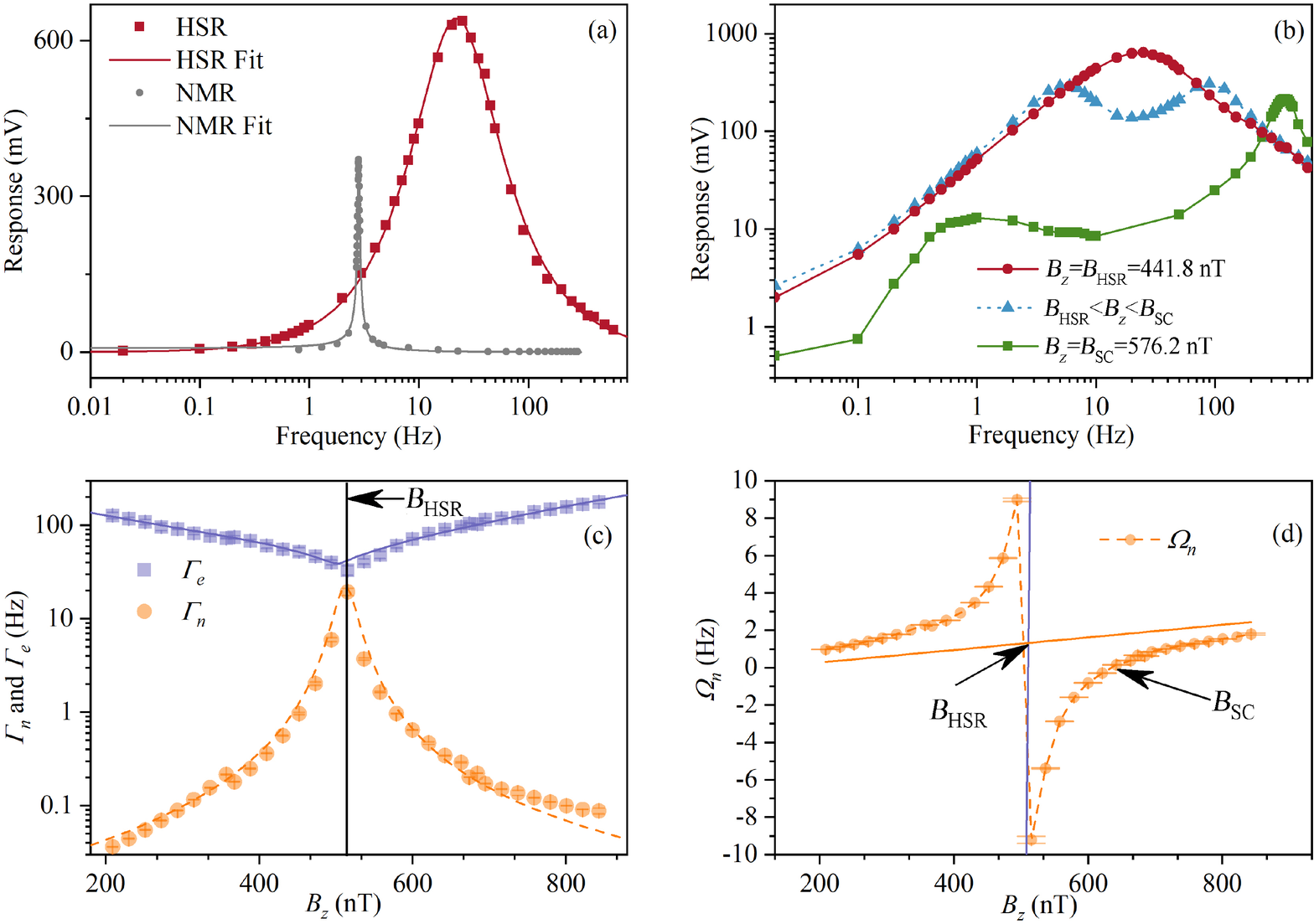}
    \caption{Proof-of-principle demonstration of a resonantly-coupled hybrid spin resonance. (a) The response comparison of the HSR and NMR regimes. (b) Responses to an oscillating magnetic field applied along $\hat{x}$ as a function of the oscillation frequency for three different working regimes: the SC regime with the external magnetic field $B_{z} = B_\text{SC} = 576.2$ nT, the HSR regime with $B_z = B_\text{HSR} = 441.8$ nT, and an intermediate regime between the two with $B_z =459.3$ nT. In (a) and (b), the comparisons are under the same conditions except for different $B_z$ values. (c) The relaxation rates  of the transverse spin components of alkali spins $\Gamma_{\rm e}$ and noble-gas spins, $\Gamma_{\rm n}$, as a function of $B_z$. (d) The precession frequency $\Omega_{\rm n}$ of noble-gas nuclear spins as a function of the bias field. 
    }
    \label{Fig.HSR.Demon} 
\end{figure*}

The dynamics of the spins can be characterized to two special regimes which have unique features. One is the SC regime, where we adjust the external magnetic field to $B_\text{SC}  =  - {B}^{\rm n}_z - {B}^{\rm e}_z$, which corresponds to approximately $576.2$ nT in Fig.~\ref{Fig.HSR.Demon}. As a result, it cancels the effective magnetic field from noble gas $\bf{{B}^{n}}$, and the slow changes in the magnetic field are automatically canceled by interactions between the alkali and noble-gas atoms~\cite{kornack2002dynamics,Wei:2022ggs}.

The second regime occurs when the external magnetic field $B_z$ is tuned to 
$B_\text{HSR} \simeq - {B}^{\rm n}_z$, which corresponds to approximately $441.8$ nT in Fig.~\ref{Fig.HSR.Demon}. Due to the screening effect of the noble gas field, the alkali atoms experience a much smaller magnetic field and precess slowly to match the precession frequency of the noble gas atoms. We term as HSR to show this merit. Due to the hybrid resonance and strong Fermi-contact interaction, the alkali and noble-gas spins become resonantly-coupled. The damping rate of alkali spins is slowed by the noble-gas spins, while the noble-gas spin damping is accelerated by the alkali spins, similar to the fast-damping effect seen in the ${}^3$He system\,\cite{kornack2002dynamics, shaham2022strong}. Consequently, the bandwidth of the noble-gas nuclear spins is significantly enhanced compared to typical NMR mode. In addition, the central frequency of the ${}^{21}$Ne HSR is more than an order of magnitude higher than that of the SC regime (see detailed explanations for Fig.~\ref{Fig.HSR.Demon}(b)).

Both SC and HSR regimes have been demonstrated in Ref.~\cite{kornack2002dynamics,xu2022critical} using the K-${}^3$He system. Besides, Ref.~\cite{shaham2022strong} also studied a strong coupling dynamics of K-${}^3$He for periodic exchange of the two spins. However, because the effective field from K atoms ${B}^{\rm e}_z$ is too weak, which is more than two orders of magnitude smaller than that of ${B}^{\rm n}_z$, these two regimes are nearly degenerate and cannot be well separated for the purpose of enhancing the bandwidth of the system. In the K-Rb-${}^{21}$Ne system demonstrated here, the ${B}^{\rm e}_z$ is more than two orders of magnitude larger than that of K-${}^3$He system, due to its stronger Fermi-contact enhancement factor and a higher alkali atom density. This enabled us to detect low frequency dark matter with enhanced sensitivity via the SC regime and to detect high frequency dark matter with enhanced bandwidth by the HSR mode.

A detailed description of the SC mode can be found in \cite{Wei:2022ggs}, and we now describe the key features of the HSR regime in the ${}^{21}$Ne system. We experimentally demonstrate the HSR regime as shown in Fig.\,\ref{Fig.HSR.Demon}. The responses of the alkali-noble-gas spins to an oscillating magnetic field are shown in Fig.\,\ref{Fig.HSR.Demon}(a) for the traditional NMR regime (gray) and  the HSR regime (red). In the NMR case, the bandwidth is approximately 0.01 Hz, which is the typical bandwidth of NMR spin magnetometers \cite{jiang2021search, xu2024constraining}. By operating the noble-gas and alkali spin ensembles in the HSR regime, the response to the magnetic field is broadened to a full-width half-maximum (FWHM) bandwidth of 36.65 Hz centering around 21.99 Hz,  which represent three orders of magnitude improvement compared with that of the NMR regime\,\cite{jiang2021search}.

In Fig.\,\ref{Fig.HSR.Demon}(b), when setting the bias field $B_z$ to the SC point $B_\text{SC}$ (green), there are two separate peaks in the frequency response. The low-frequency peak originates from noble-gas nuclear spins, while the high-frequency peak is from alkali spins. For the bias field $B_z$ between the SC point $B_\text{SC}$ and the HSR point $B_\text{HSR}$ (blue), the response peaks of noble-gas nuclear spins and alkali spins move closer together. Finally, when the bias field is set to the HSR point $B_\text{HSR}$ (red), the two peaks merge and form a broadened HSR peak. The central peak frequencies of the ${}^{21}$Ne system in the HSR and SC regimes are approximately 22 Hz and approximately 0.9 Hz, respectively, separated by more than an order of magnitude, which is a clear distinction from the ${}^{3}$He system.

To present the changes in bandwidth of alkali  spins and noble-gas spins, we measure the relaxation rate (damping rate) of the transverse spin component of two spin species respectively. As shown in Fig.\,\ref{Fig.HSR.Demon}(c), the damping rate $\Gamma_{\rm e}$ of alkali electron spins decreases as the bias field $B_z$ approaches the HSR point $B_\text{HSR}$, while the damping rate $\Gamma_{\rm n}$ of noble-gas nuclear spins increases from approximately 0.04 Hz to about 20 Hz.  This result intuitively presents that the relaxation rate of the alkali electron spins is slowed down by resonant coupling with noble-gas nuclear spins, while the noble-gas nuclear spins are sped up. Therefore, the hybrid response peak is no longer contributed by two spin species separately. Rather, the peak is due to the  resonantly-coupled hybrid spin ensembles. The bandwidth of the HSR is a hybrid combination of $\Gamma_{\rm e}$, $\Gamma_{\rm n}$, ${{B}^{\rm e}_z}$, and ${{B}^{\rm n}_z}$, and also depends on the bias field $B_z$ (see Materials and Methods).

As shown in Fig.\,\ref{Fig.HSR.Demon}(d), when working away from the HSR point, the noble-gas precession frequency $\Omega_{\rm n}$ gradually approaches the asymptotic frequency (orange solid line) $\Omega_{\rm n}=\gamma_{\rm n} (B_{z}+{B}^{\rm e}_{z})$, which is the Larmor precession  of uncoupled noble-gas spins. 
However, when working around the HSR point, the noble-gas's precession rate approaches the asymptotic frequency (purple solid line) $\Omega_{\rm e}=\gamma_{\rm e} (B_{z}+{B}^{\rm n}_{z})$, which is the precession frequency of uncoupled alkali atoms.  In this region, because of the resonant coupling between alkali and noble gas atoms, the precession of noble-gas is dominated by the alkali atoms, and its precession frequency reaches the maximum. The hybrid responses of two types of atomic spins provide a new way to manipulate the gyromagnetic ratio of different atoms.
\\

\noindent \textbf{{Experimental Sensitivity}}

\begin{figure}[htb]
\centering
\includegraphics[width=0.42 \textwidth]{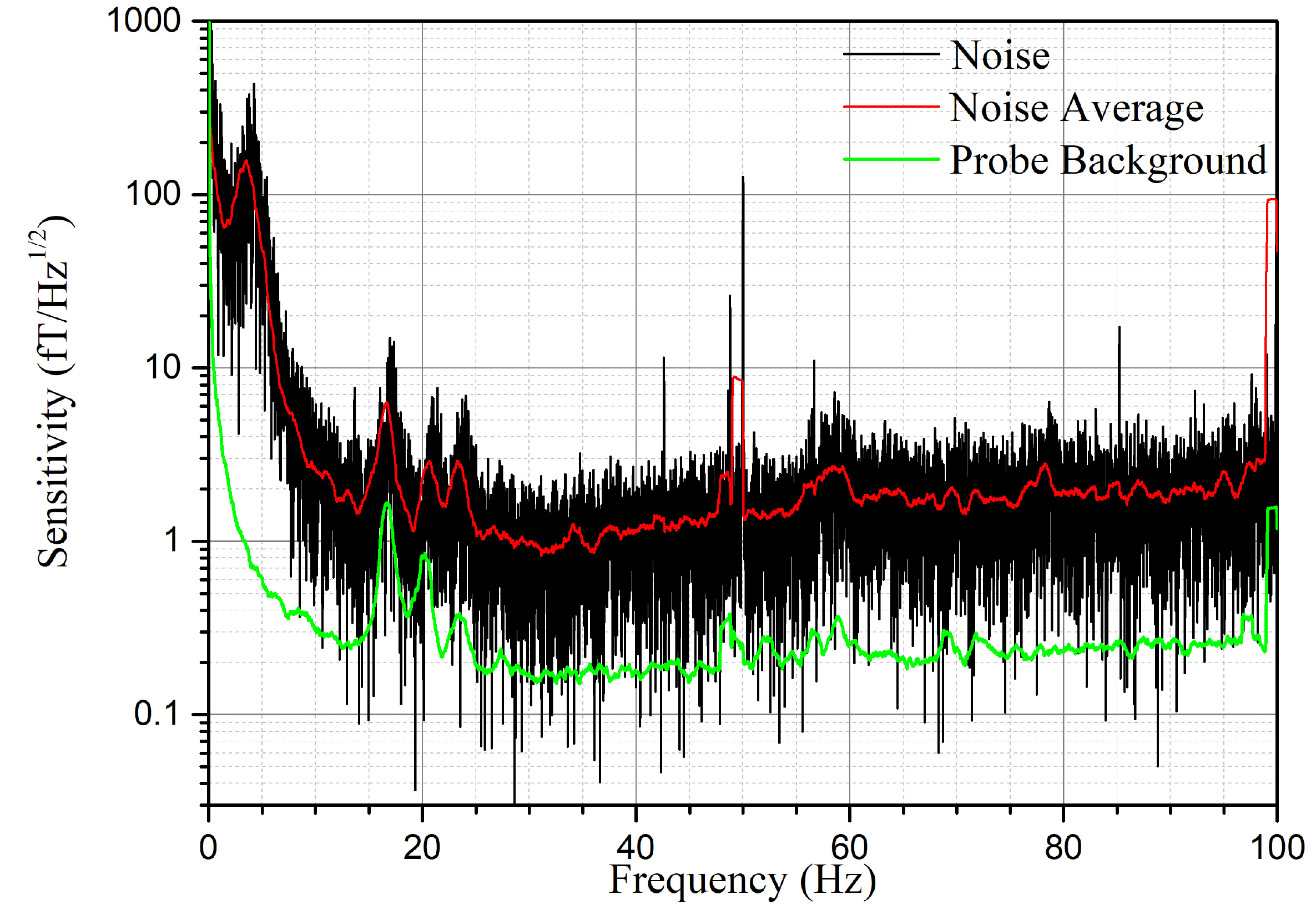}
\caption{The sensitivity at the hybrid spin resonance to a $\hat{y}$-directed magnetic field is  0.78\,fT/Hz$^{1/2}$ from 28 to 32\,Hz, placing the system among the most sensitive sensors using alkali-metal spins to measure the precession of noble-gas nuclear spins. The noise peaks around 6\,Hz, and from 17 to 23\,Hz are due to the vibrations, which is verified with a seismometer. The noise peak at 50\,Hz is the power-line noise. The background noise of the probe with pump light blocked is much smaller than the total noise level.}
\label{Fig.sensivity} 
\end{figure}

We now discuss the magnetic and dark-matter sensitivity of the HSR regime. The sensitivity of the setup is calibrated by applying oscillating magnetic fields generated with a magnetic coil. The spectrum of the setup, shown in Fig.\,\ref{Fig.sensivity}, demonstrates a sensitivity of 0.78 fT/Hz$^{1/2}$ from 28 to 32 Hz, which is remarkably high for comagnetometers. The magnetic noise of the inner Mn-Zn ferrite shield in the relevant frequency range is estimated to be 2.5\,$f^{-1/2}$\,fT ($f$ is the frequency in Hz)~\cite{kornack2007low}, using the measured relative permeability and geometrical parameters of the shield. Details about vibration and magnetic noise can be found in the supplementary information in Ref.\,\cite{wei2022constraints}. This means that the estimated magnetic noise from the shield is close to the total noise level. The background noise of the probe with pump light blocked is about 0.2\,fT/Hz$^{1/2}$ from  28 to 32\,Hz, which is significantly smaller than the total noise. The spin-projection noise of alkali spins is calculated to be about 0.09\,fT/Hz$^{1/2}$ \cite{kornack2005nuclear}.

In HSR regime, the relationship between the response to ultralight dark matter field $b^{\rm n}_x$ coupling with noble-gas nuclear spins $S^{\rm e}_x = K_{b^{\rm n}_x} b^{\rm n}_x$ and the response to magnetic field $S^{\rm e}_x = K_{B_y} B_y$ is derived in Materials and Methods in details. The two response factors are related as,
\begin{align}
     K_{B_y}=K_{b^{\rm n}_x} \omega / {\omega}^{\rm n}_z ,
\end{align}
where ${\omega}^{\rm n}_z =  \gamma_n {B}^{\rm n}_z$ and $\omega$ is the angular frequency of the signal $b^{\rm n}_x$. Therefore, we can use the normal magnetic field to calibrate the response to the pseudomagnetic field $b^{\rm n}_x$. \\

\noindent \textbf{{Axion Dark Matter Searches}}

The gradient axion field can couple with the nucleon magnetic dipole moment and can be viewed as a pseudomagnetic field, \(\mathbf{b}_a\). By measuring its effects on the nuclear spins, separate from the normal magnetic field, we can identify its existence and constrain its magnitude.
The gradient of the axion dark matter field in a volume $V$ can be written as
	\begin{align}
		\bm{\nabla} a(x) = \sum_{\bf{p}} \sqrt{\frac{2 N_{\bf{p}}}{V \omega_{\bf{p}}}} \cos(\omega_{\bf{p}} t - \bf{p } \cdot \bf{x}+ \phi_{\bf{p}} ) \bf{p}\,,
	\end{align}
where the sum runs over all momentum modes $\bf{p}$, $\phi_{\bf{p}}$ is the random initial phase related to the mode $\bf{p}$ modeled as a uniform variable in the $[0,\,2\pi]$ interval, $N_{\bf{p}} = \rho_{\rm DM} V f(\bf{p})$$
(\Delta p)^3/\omega_{\bf{p}}$ is the mean occupation number of mode $\bf{p}$. The function $f(\bf{p})$ is the Maxwell-Boltzmann velocity distribution for DM in the Standard Halo Model \cite{bovy2009galactic} and is normalized to $\int d\mathbf{p} ~f(\mathbf{p}) = 1$, and $\rho_{\rm DM} = 0.4\,{\rm GeV}\cdot {\rm cm}^{-3}$ is an estimate of local DM energy density\,\cite{de2021dark}. Possibilities of local enhancement of both the density and coherence time of the dark matter field have been considered (see, for example, \cite{banerjee2020searching}).
Nevertheless, we present our analysis based on the Standard Halo Model. 

The sum over all $\bf{p}$ modes leads to a stochastic pattern \cite{foster2018revealing, centers2021stochastic, lisanti2021stochastic, gramolin2022spectral, lee2022laboratory}, which can be noticed for an experiment duration longer than the characteristic coherence time $\tau_a \approx m_a^{-1} \sigma_{v}^{-2}\hbar$\,\cite{derevianko2018detecting} with the DM velocity dispersion $\sigma_v = 220/\sqrt{2}~{\rm km}\cdot{\rm s}^{-1}$. 
We account for the DM stochastic effects in the analysis following the frequency-domain likelihood-based formalism of Ref.\,\cite{lee2022laboratory}.	

For high-frequency measurements, we conducted a 209-hour measurement with an $86\%$ duty cycle (Dataset 1) and a 4-hour continuous measurement with a nearly identical setup deployed in an underground laboratory (Dataset 2), which provided better suppression of vibrations and power-line interference. Both measurements were conducted at a fixed bias magnetic field in the HSR regime. For low-frequency measurements, we employed the SC regime and collected 146 hours of data (Dataset 3).
We used the log-likelihood ratio (LLR) test to analyze the datasets, obtain exclusion limits for the magnitude of the pseudomagnetic field, and convert these to limits for axion-nucleon couplings. The calculations closely follow those in Ref.~\cite{lee2022laboratory}, with details provided in Materials and Methods.

 \begin{figure}[]
		\centering
		\includegraphics[width= 0.98  \linewidth]{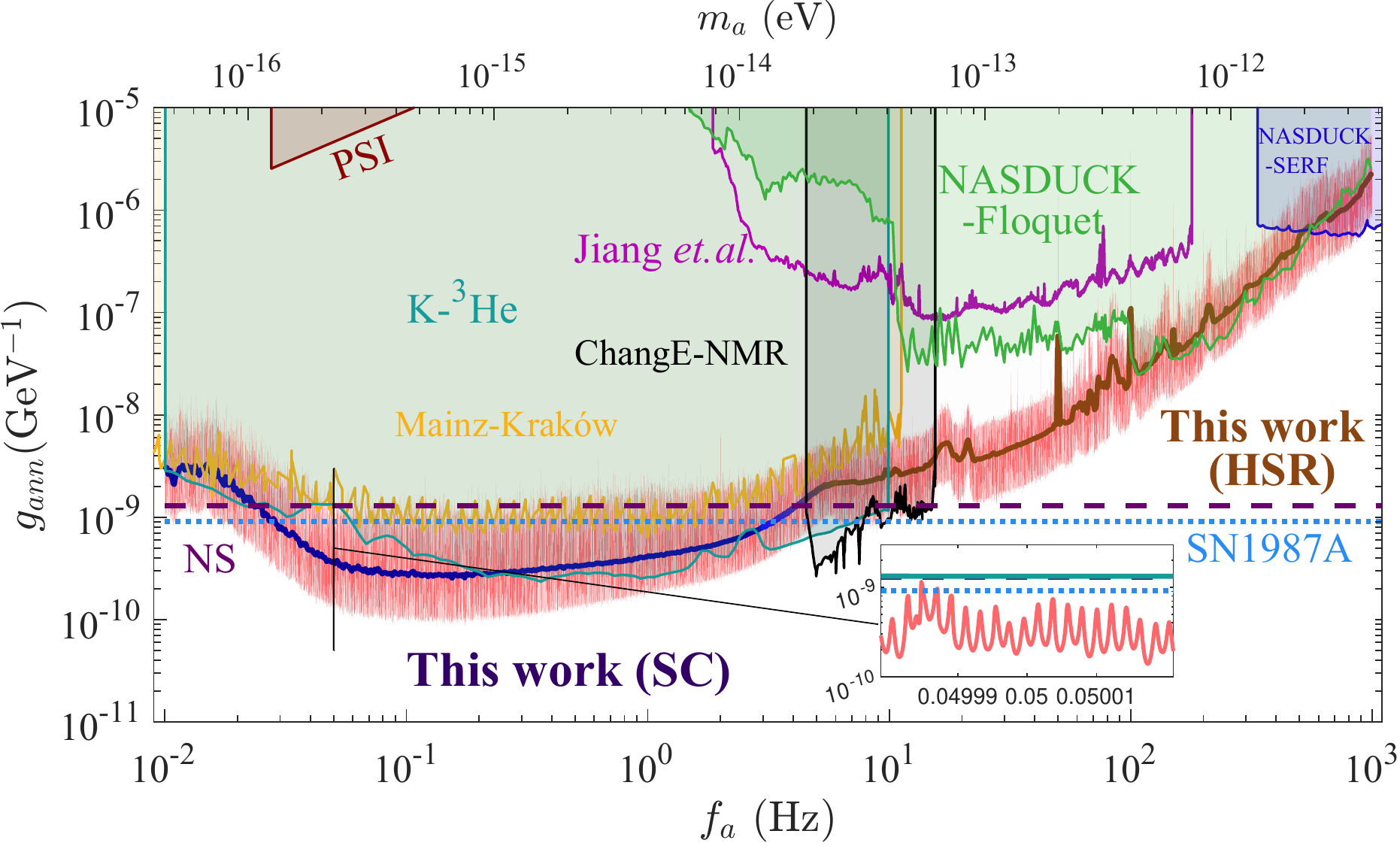}
		\includegraphics[width= 0.98 \linewidth]{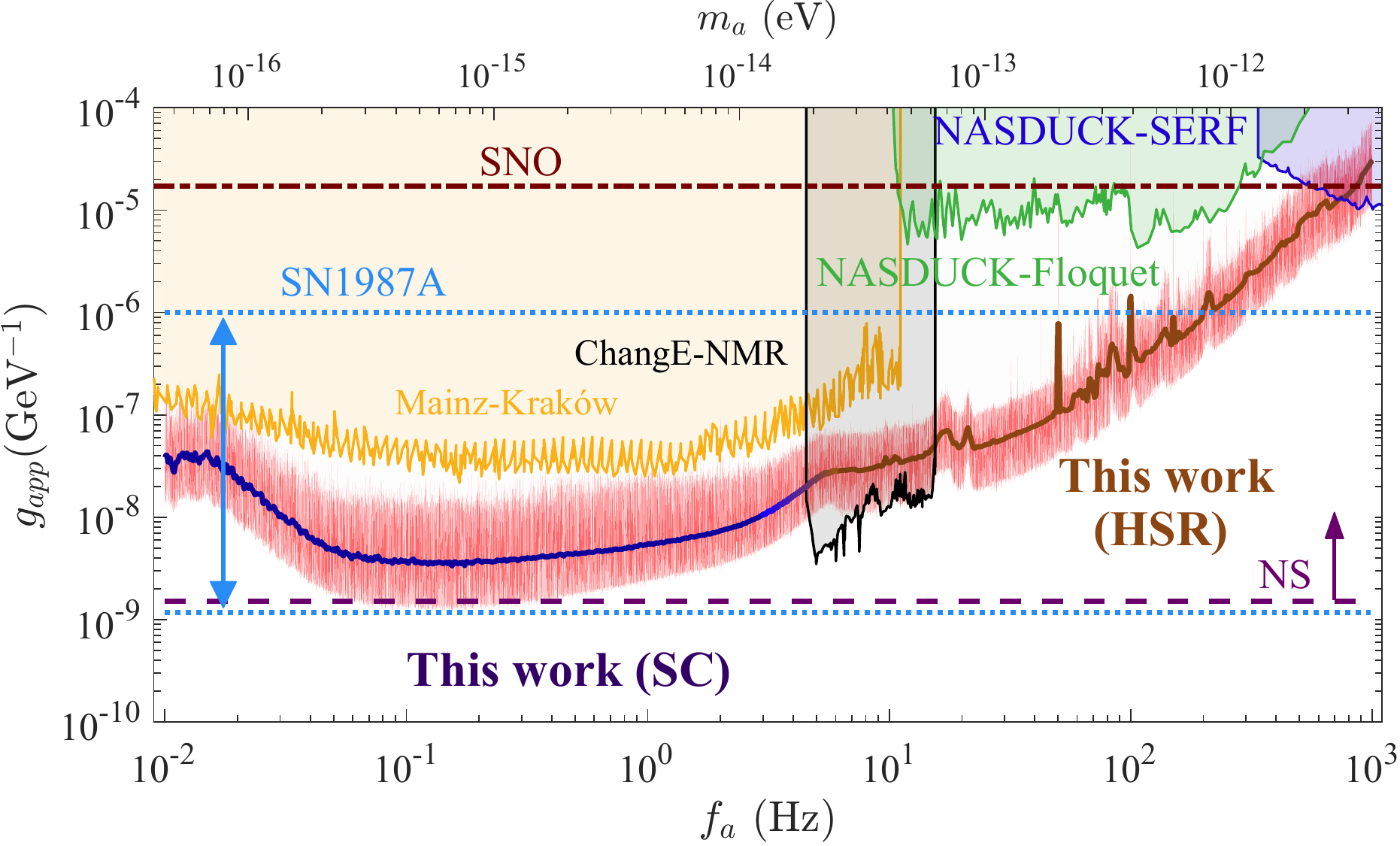} 
\caption{
The $95\%$ C.L. upper limits on the axion-neutron coupling $g_{ann}$ and axion-proton coupling $g_{app}$ from both the SC and HSR measurements are shown as red lines.
The full data of the limits cannot be shown in the figure, but are tabulated in~\cite{axiondata}. To guide the readers' eye, for each $f_a$, the couplings $g_{ann}$ and $g_{app}$ are averaged in a bin from $0.99 f_a$ to $1.01 f_a$, following \cite{bloch2022new}. 
The averaged limits are shown in dark blue for SC data (labeled as ``This work (SC)", covering the range $[0.01, 6]$ Hz) and in brown for HSR data (labeled as ``This work (HSR)", covering the range $[3, 1000]$ Hz).
For the $[3, 6]$\,Hz range, we choose the stronger limits from the SC and HSR results.
We also show other terrestrial limits from our previous results with bandwidth-enhanced NMR experiment \cite{xu2024constraining} (ChangE-NMR) as well as that of Jiang et al. \cite{jiang2021search} with a conversion from nucleus coupling to nucleon coupling implemented, ${\rm K}$-$\ce{^{3}He}$ comagnetometer \cite{lee2022laboratory}, NASDUCK-Floquet \cite{bloch2022new}, NASDUCK-SERF \cite{bloch2023constraints}, Mainz-Krak$\Acute{\text{o}}$w \cite{gavilan2024searching}, CASPEr-ZULF \cite{wu2019search} and PSI \cite{abel2023search}. The astrophysical limits from neutron star (NS) cooling \cite{buschmann2022upper}, supernova SN1987A \cite{carenza2019improved} and solar axion at SNO \cite{bhusal2021searching} are shown as horizontal lines respectively.}
		\label{fig:axion-limits} 
\end{figure} 

In Fig.\,\ref{fig:axion-limits}, we set exclusion limits (95\% C.L.) in red for axion-neutron $g_{ann}$ and axion-proton $g_{app}$ couplings. 
For the axion-neutron coupling $g_{ann}$, the HSR limits reach down to $\mathcal{O}(3\times 10^{-9}\,{\rm GeV}^{-1})$, close to the astrophysics limits and improve on the results from NASDUCK-Floquet \cite{bloch2022new} by 1-2 orders of magnitude. Note that the work \cite{jiang2021search} also reported a series of ``needle'' exclusions in the frequency range of 1\,Hz to 200\,Hz with a typical width of 35.8\,mHz. When comparing these results, it is essential to consider the stochastic correction and the neutron polarization correction, $\xi_{\rm n}$ in $^{129}$Xe. 
In order to attain continuous exclusion, the study presented in \cite{jiang2021search} necessitates an accumulation period of about 3 years, while our data takes a significantly shorter duration of just one week.
The SC limits are comparable to the K-$^3{\rm He}$ results~\cite{lee2022laboratory}, while providing slightly better constraints for the frequency range of [$2\times 10^{-2}$, $2\times 10^{-1}$]\,Hz. This improved sensitivity is due to the smaller gyromagnetic ratio of $^{21}{\rm Ne}$ compared to $^3{\rm He}$, despite the  shorter measurement time 146\,hr comparing with the 40-day duration of K-$^{3}$He results \cite{lee2022laboratory}. Our results show slightly weaker sensitivity in the frequency range of 4.5 to 15.5 Hz compared to our previous bandwidth-enhanced NMR magnetometer work~\cite{xu2024constraining}, but they cover a significantly larger bandwidth in terms of axion mass.
Our laboratory limits for $g_{ann}$ are stronger than the astrophysical limits based on the emission of neutron stars \cite{buschmann2022upper} and supernova SN1987A \cite{carenza2019improved} for low $f_a$ and are comparable at intermediate frequencies $\sim \mathcal{O}(10)$\,Hz. It is important to note that astrophysical limits are subject to various uncertainties, such as density-dependent coupling, unknown heating mechanisms for neutron stars \cite{beznogov2018constraints, buschmann2022upper, di2022stellar}, and axion production, scattering, and absorption in dense plasma environments, as well as the model dependence of the collapse mechanism for supernovae \cite{chang2018supernova, carenza2019improved, bar2020there}. 

For the axion-proton coupling $g_{app}$, we have achieved the most stringent terrestrial constraints across an extensive frequency range of $[2\times 10^{-2}, \, 5]$\,Hz and $[16, \, 7\times 10^{2}]$\,Hz using the SC and HSR data, complementary to our previous enhanced NMR experiment~\cite{xu2024constraining}. In principle, the proton's fraction of spin in $^3$He can be taken into account, then the K-$^3$He system can also provide a strong limit on $g_{app}$ in the corresponding frequency range. In Fig.\,\ref{fig:axion-limits}, we compared our results with previous results. The precise proton fraction $\xi_{\rm p}^{\rm Xe}$ within the Xenon nuclear spin is associated with unbounded uncertainties~\cite{bloch2022new}. Therefore, we select the conservative result among the four models presented in NASDUCK-Floquet~\cite{bloch2022new}.

Aside from setting limits on axion couplings, a post-analysis is carried out (see Materials and Methods) to test the significance of the best-fit signal compared to the background-only model and to statistically assess any data points that exceed the $5\sigma$ threshold. With the look-elsewhere effect taken into account, we use ${\rm LLR} > 52.1$ for a one-sided global significance of a $5\sigma$ test~\cite{lee2022laboratory}. Combining the two HSR data sets, we found that only 62 out of 6.8 million tested axion masses surpassed the $5\sigma$ confidence level, a small fraction of the total data set. No further $\geq 5\sigma$ candidates were identified in Dataset 3 (SC). This indicates that the majority of the data are consistent with the white noise background assumption. Nonetheless, a few peaks ``survived" and it is possible that they could originate from axion DM. Consequently, we subjected the 62 candidates to three additional tests, as detailed in Materials and Methods, which led to the exclusion of these candidates as being a dark matter origin. 
Note that the other results, for example NASDUCK-SERF \cite{bloch2023constraints} have also seen numerous `candidate' peaks in the data that are likely due to systematic effects.
Though, the candidate at a frequency of $f_a = 46.704996$ Hz is marginally excluded at the 95\% confidence level.
We emphasize that quantitative exclusions may be less reliable due to the potential violation of the white noise assumption resulting from the presence of systematic noises. It would be useful to revisit these points in future studies where spurious peaks may arise at different locations with independent setups, thus these regions can be covered. 
\\

\noindent \textbf{\large{DISCUSSION}}

We demonstrate a new HSR regime that can be used for a broadband search for new physics such as dark matter.  The embedded alkali SERF  magnetometer enables high sensitivity to magnetic fields, and the HSR coupling broadens the bandwidth. This work paves a way to enhance the bandwidth and the sensitivity simultaneously, contrary to those improving the bandwidth at the cost of sensitivity degradation. We operated the alkali-noble-gas spin ensembles in SC and HSR working regimes, searching for dark matter in the frequency range from 0.01\,Hz to 1\,kHz. We found a number of candidates but rejected them based on their non-compliance with the expected properties of DM signals.  We obtained limits that surpass previous laboratory results and, within a certain axion mass range, even exceed those from astrophysics with strong model assumptions.
In addition, for the axion-proton coupling $g_{app}$, we achieved the most stringent terrestrial constraints for the frequency range $[2\times 10^{-2}, \, 5]$\,Hz and$[16, \, 7\times 10^{2}]$\,Hz.

The current dominant source of noise is magnetic noise from the shielding material. To improve the performance of the comagnetometer, additional active magnetic noise compensation loops can be utilized. Using better shielding material (for example, superconducting) or a bigger shielded room can also help since magnetic noise reduces with the shield size. In scenarios requiring the measurement of low-frequency signals, such as low-frequency dark matter detection and inertial rotation measurements, the SC mode can be utilized due to its effective suppression of low-frequency magnetic noise. Further improvements can be achieved by reducing the effective magnetic field gradient and enhancing the effective magnetic field from nuclear spins, thereby increasing the noise suppression capability of the SC mode. In future work, the HSR/SC magnetometer can be utilized to search for new physics including exotic spin-dependent forces\,\cite{wei2022constraints}. A network of HSR/SC magnetometers can be applied to search for topological defect dark matter\,\cite{afach2024can}, and can work in the intensity interferometry mode to detect dark matter in a much higher frequency range above the magnetometer bandwidth~\cite{masia2022intensity}. The broadband search can also be applied in other fields that need high bandwidth and ultrahigh sensitivity, for instance the measurement of brain signal for magnetoencephalogram\,\cite{boto2018moving}, as well as the measurement of magnetism of ancient rocks to study global climate change history\,\cite{tauxe2010essentials}. 
\\

\vspace{0.1 cm}

\noindent \textbf{\large MATERIALS AND METHODS}

\noindent \textbf{Detailed Experimental Configurations}

\begin{figure}[htb]
    \centering
	\includegraphics[width= 0.48 \textwidth]{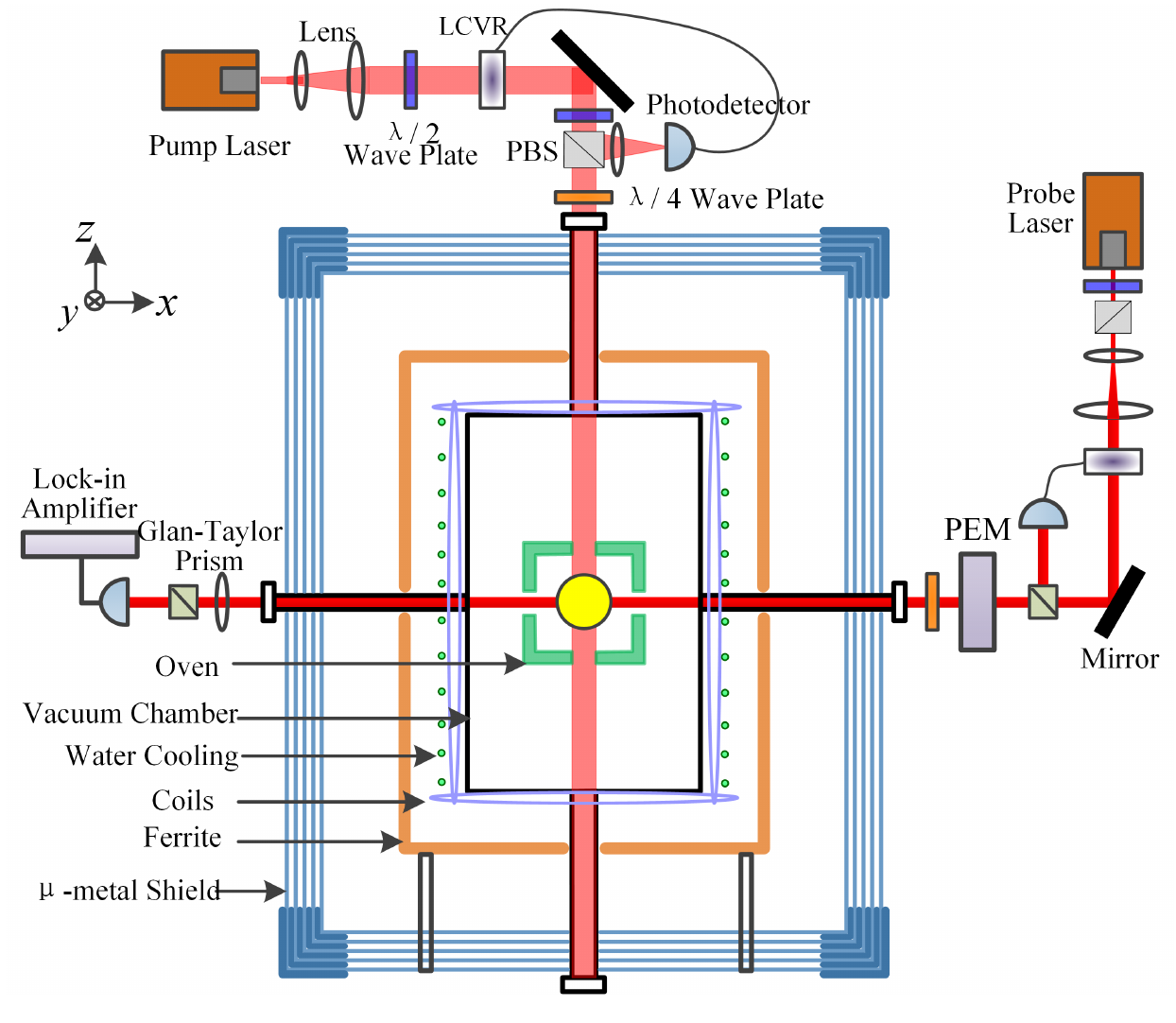}
    \caption{
    The illustration of the experiment setup. At the center of the device is a glass cell containing a droplet of mixed K-Rb alkali metal, 3\,amg $^{21}$Ne, 0.066\,amg N$_2$, is installed in an oven, which is heated with an AC electric heater. The oven is enclosed in a plastic (PEEK) vacuum chamber to reduce the heat dissipation and air convection. The water cooled vacuum chamber also acts as the frame for triaxial magnetic coils. The vacuum chamber is enclosed in five layers of $\mu$-metal magnetic shielding, and an inner Mn-Zn ferrite shield used to minimize the Johnson-current magnetic noise.  
   The probe light is polarization modulated with a photoelectric modulator (PEM) and demodulated with a lock-in amplifier to avoid the low-frequency noise. The pump and probe lasers are frequency- and amplitude-stabilized, as described in Supplementary Information of Ref.\,\cite{wei2022constraints}. PBS stands for polarization beam splitter.
    }
    \label{Fig1.setup} 
\end{figure}

The experimental setup is demonstrated in Fig.~\ref{Fig1.setup}. Within a 11.4 mm diameter spherical cell, a small droplet of K and Rb alkali metal atoms is combined with about 3 amg $^{21}$Ne (70$\%$ isotope enriched) and 0.066 amg N$_2$. The density ratio of K to Rb is set to approximately \(1/55\) to enhance polarization efficiency, improve uniformity, and increase sensitivity~\cite{lancor2011polarization, babcock2003hybrid, Wei:2022ggs}. 

To fabricate a hybrid alkali-metal cell with a designed density ratio, a specialized process is required beyond standard cell preparation. First, the alkali-metal mixture is carefully weighed in an anaerobic environment to maintain the desired ratio. Next, the mixture is connected to a gas chamber fabrication device. It is introduced into the gas chamber using a controlled flame-sealing technique. Finally, after the glass cell is sintered down, careful testing of density ratio and selection of glass cell are necessary.

To maintain a stable temperature while reducing noise, the cell is heated to around 473 K using electric heaters. The AC electric heater is located inside a PEEK vacuum chamber with water-cooling. This design reduces heat dissipation and minimizes the impact of air convection on the pump and probe lights. Additionally, the water-cooled chamber helps lower the temperature of surrounding magnetic shields, thereby improving the magnetic noise performance. The current through the heaters is modulated at a high frequency of 200 kHz to minimize low-frequency magnetic noise from heating. In addition, a double-layer printed heating coil is employed to cancel out the magnetic noise from each heating coil. A homemade heating closed-loop control system ensures temperature stability.

The atomic cell is enlcosed in a ferrite shield and a five-layer $\mu$-metal magnetic shield to provide an ultralow magnetic noise condition. The magnetic noise of ferrite shield itself is optimized by comparing the performance of different material composition. The optimized ferrite shield features a low intrinsic magnetic noise at the level of sub-fT/Hz$^{1/2}$. Inside the ferrite shield, three-axis coils wrapped  around the vacuum chamber further reduce the residual magnetic field due to the shield and facilitate spin ensemble manipulation.

Hybrid spin-exchange optical pumping is employed. A circularly polarized pump light along the \textit{z}-axis is locked to the D1-line frequency of K atoms and utilized to polarize the optically-thin K atoms, with a light intensity of approximately 450 mW/cm$^2$. The optically-thick Rb atoms undergo spin polarization through spin-exchange collisions with K atoms, thereby reducing strong light absorption by the optically-thick alkali atoms and improving the spin polarization homogeneity of alkali atoms. Furthermore, the $^{21}$Ne nuclear spins are polarized through spin-exchange collisions with alkali atoms, primarily Rb atoms.

A linearly polarized probe light along the \textit{x} axis measures the transverse component of electronic spin polarization of Rb atoms based on the optical rotation. The frequency of probe light is tuned to the low-frequency side of D1-line of Rb by about 240\,GHz. Glan-Taylor prisms with an extinction ratio of about 100000:1 are used as polarizer and analyzer in the probe light path. The probe light, modulated by a photoelastic modulator at 50 kHz, isolates low-frequency noise. Pump and probe lasers, closed-loop stabilized with liquid crystal variable retarders (LCVR), consist of a laser diode from Eagleyard and a homemade control system. Additional key details of the experimental setup include the electronic spin polarization of approximately 0.5, the electronic spin relaxation rate of about $ 3380 \, \text{rad/s} $, the nuclear spin polarization of approximately 0.07 and the nuclear spin relaxaiton rate of about 0.06\, $\text{rad/s} $ measured in the SC regime.
\\

\noindent \textbf{Dynamics of Hybrid Spin Resonance}

The collective polarizations of alkali and noble-gas spins are described by ${\bf{ P^{e}}}={\bf{ \left< S_{e} \right>}}/S_\text{e}$ and ${\bf{ P^{n}}}={\bf{\left<K_{n}\right>}}/K_\text{n}$, respectively. In this context, $\bf{S_e}$ represents the valence electron spin operator (with $S_\text{e}=1/2$) of the alkali atom, while $\bf{K_n}$ denotes the nuclear spin of the noble-gas atoms ($K=1/2$ for $^{3}$He and $K=3/2$ for $^{21}$Ne). The dynamics of these spins, which occupy the same volume (i.e., the glass cell), can be characterized by the Bloch equations, which couple the alkali electron spin polarization ${\bf{P^{\rm e}}}$ with the noble-gas nuclear spin polarization ${\bf{P^{\rm n}}}$ \cite{kornack2005nuclear}.
\setcounter{equation}{0}
\renewcommand{\theequation}{A.\arabic{equation}}

\begin{eqnarray}\label{eq.Bloch}
   \frac{{\partial  {{\bf{P^{\rm e}}}} }}{{\partial t}} =&&  \frac{\gamma _\text{e}}{Q}\left({\bf{B}} + {\bf{{B}^{\rm n}}} + {\bf{b^{\rm e}}} -{\Omega}\frac{Q }{\gamma _\text{e} } +{{\bf{L^{\rm e}}}} \right) \times  {{\bf{P^{\rm e}}}} \nonumber \\ 
   && +\frac{R_\text{p} S^\text{p}_{z} +R^{\rm{ne}}_{\rm{se}}{\bf{P^{\rm n}}}}{Q} -\frac{\{{R^\text{e}_1, R^\text{e}_2, R^\text{e}_2}\}}{Q}{\bf{P^{\rm e}}} +D_\text{e} \nabla^2 {{\bf{P^{\rm e}}}} ,\nonumber \\
   \frac{{\partial  {{{\bf{P^{\rm n}}}}} }}{\partial {t}} =&&  \gamma _\text{n} \left({\bf{B}} + {\bf{{{B}}^{\rm e}}}  + {\bf{b^{\rm n}}} - \frac{\Omega}{\gamma _\text{n}} \right) \times {\bf{P^{\rm n}}} \nonumber \\
   && + R^{\rm{en}}_{\rm{se}} {\bf{P^{\rm e}}}  - {\{  R^\text{n}_1, R_2^\text{n},  R_2^\text{n}  \}} {{\bf{P^{\rm n}}}}+ D_\text{n} \nabla^2 {{\bf{P^{\rm n}}}}.
\end{eqnarray}

The first equation describes the dynamics of alkali electron spins, while the second equation describes the dynamics of noble-gas nuclear spins. Each equation consists of four terms on the right-hand side. The first term describes the spin precession of the spin ensemble. Alkali electron spins precess under the sum of a classical magnetic field ${\bf{B}}$, an effective field ${\bf{{B}^{\rm n}}}$ from the noble-gas spins, an exotic field coupling to the electron spins due to ultralight dark matter ${\bf{b^{\rm e}}}$, an inertial rotation $\Omega$, and an AC Stark light shift ${\bf{L^{\rm e}}}$. Similarly, noble-gas spins precess under the sum of ${\bf{B}}$, an effective field ${\bf{{B}^{\rm e}}}$ from the alkali spins, an exotic field coupling to the nuclear spins due to ultralight dark matter ${\bf{b^{\rm n}}}$, and $\Omega$. Here, $\gamma_\text{e}$ and $\gamma_\text{n}$ represent the gyromagnetic ratios of alkali electrons and noble-gas nuclei, respectively. $Q$ is the slowing-down factor of alkali atoms considering the hyperfine interaction from its nuclear spin.

The second term of each equation describes the spin polarization of alkali and noble-gas spins. Alkali spins are optically polarized using circularly polarized pump light at a pumping rate $R_\text{p}$, with photon spin $S^\text{p}_{z} \approx 1$. Additionally, alkali spins are polarized via spin-exchange collisions with noble-gas spins at a rate $R^{\rm{ne}}_{\rm{se}}$, which is negligible compared to the pumping rate $R_\text{p}$. Conversely, noble-gas spins are polarized via spin-exchange collisions with alkali spins at a rate $R^{\rm{en}}_{\rm{se}}$.

The third term of each equation describes the spin relaxation of the spin ensemble. The longitudinal and transverse spin components of alkali spins relax at rates $R^\text{e}_1$ and $R^\text{e}_2$, respectively, while the longitudinal and transverse spin components of noble-gas spins relax at rates $R^\text{n}_1$ and $R^\text{n}_2$, respectively.
The final term of each equation characterizes the diffusion of each spin species, where $D_\text{e}$ and $D_\text{n}$ represent the diffusion constants of alkali and noble-gas atoms, respectively.

When a small transverse excitation is introduced, the longitudinal components of spin polarizations, $P^\text{e}_{z}$ and $P^\text{n}_{z}$, remain nearly unchanged and are equal to their respective equilibrium values, $P^\text{e}_\text{z0} = R_\text{p}S^\text{p}_{z}/R^\text{e}_1$ and $P^\text{n}_\text{z0}=P^\text{e}_\text{z0}R^{\rm{en}}_{\rm{se}}/R^\text{n}_1$. Therefore, the longitudinal components can be factored out of the coupled Bloch equations. To reduce the number of equations, the transverse components of alkali and noble-gas spins can be expressed as $ P_{\bot}^\text{e} = P_{x}^\text{e} + {\rm i} P_{y}^\text{e} $ and $ P_\bot^\text{n} = P_{x}^\text{n} + {\rm i} P_{y}^\text{n} $, respectively.

The dynamic response of the coupled alkali and noble-gas spin ensembles can be determined by solving the coupled Bloch equations. In this experiment, we measure the transverse component of alkali spins, $P^\text{e}_{x}(t)$, which can be expressed as:
\begin{equation} 
P_{x}^\text{e}(t)  = \text{Re}[P_1 \text{e}^{-(\Gamma_\text{e} + {\rm i} \Omega_\text{e})t} + P_2 \text{e}^{-(\Gamma_\text{n} + {\rm i} \Omega_\text{n})t}] +P_0\,,
 \label{eq.dynamic}
\end{equation}

\noindent where the two exponential terms correspond to the alkali and noble-gas spins, respectively. The parameters $\Gamma_\text{e}$ and $\Omega_\text{e}$ represent the decay rate and precession frequency of the alkali spins, while $\Gamma_\text{n}$ and $\Omega_\text{n}$ correspond to the noble-gas spins. The coefficients $P_1$, $P_2$, and $P_0$ are dependent on the specific working conditions. The decay rate and precession frequency of each spin species are interconnected with those of the other species, and can be expressed as:
\begin{align} 
\Gamma_\text{e}  & = \frac{ R_2^\text{e} /Q + R_2^\text{n} }{2} + \frac{1}{2\sqrt 2 }\sqrt {\sqrt {a^2  + b^2 }  + a} , \nonumber\\
 \Omega_\text{e}  & = - \frac{ \omega^\text{e}_{z} + \omega^\text{n}_{z} }{2} + \frac{1}{2\sqrt 2 } \text{Sign}(b) \sqrt {\sqrt {a^2  + b^2 }  - a} , \nonumber \\
 \Gamma_\text{n}  & = \frac{ R_2^\text{e} /Q + R_2^\text{n} }{2} - \frac{1}{2\sqrt 2 }\sqrt {\sqrt {a^2  + b^2 }  + a} , \nonumber \\
 \Omega_\text{n}  & = - \frac{ \omega^\text{e}_{z} + \omega^\text{n}_{z} }{2} - \frac{1}{2\sqrt 2 } \text{Sign}(b) \sqrt {\sqrt {a^2  + b^2 }  - a} \,,
 \label{eq.decay_prece}
\end{align}

\noindent where the parameters $a$ and $b$ are given by
\begin{equation} 
\begin{split}
a = {( R_2^\text{e} /Q  - R_2^\text{n} )^2} - {(\omega^\text{e}_{z}  - \omega^\text{n}_{z} )^2} + 4 R_{\rm{se}}^{\rm{ne}} R_{\rm{se}}^{\rm{en} }/Q  - 4\omega^{\rm{ne}}_z  \omega^{\rm{en}}_z,\\ 
b =  - 2( R_2^\text{e}/Q  - R_2^\text{n} )(\omega^\text{e}_{z}  -  \omega^\text{n}_{z} ) - 4( R_{\rm{se}}^{\rm{ne}} \omega^{\rm{en}}_z /Q   +  R_{\rm{se}}^{\rm{en}} \omega^{\rm{ne}}_z )\,,
 \label{eq.parameters_ab1}
 \end{split}
\end{equation}

\noindent where $\omega^\text{e}_{z}= \gamma_\text{e} ({B}^\text{n}_{z}+B_{z})/Q$, $\omega^\text{n}_{z}= \gamma_\text{n} ({B}^\text{e}_{z}+B_{z})$, $\omega^{\rm{ne}}_{z}= \gamma_\text{e} \lambda M^\text{n}_0 P^\text{e}_{z}/Q$, and $\omega^{\rm{en}}_{z}= \gamma_\text{n} \lambda M^\text{e}_0 P^\text{n}_{z}$. 

The coupling between alkali spins and noble-gas spins is determined by the dominant field $B_{z}$. When $B_{z}$ is much greater than ${B}^\text{n}_{z}$, the dynamics of alkali spins and noble-gas spins are decoupled. The alkali spins decay and precess based on their own relaxation rate $R^\text{e}_2/Q$ and precession frequency $\gamma_\text{e} (B_{z}+{B}^\text{n}_{z})/Q$, while noble-gas spins exhibit similar behavior, with a relaxation rate $R^\text{n}_2$ and a precession frequency of $\gamma_\text{n} (B_{z}+{B}^\text{e}_{z})$.  However, at the HSR point, alkali and noble-gas spins become resonantly-coupled. This
leads to hybridization between noble-gas spins and alkali spins, resulting in the maximum decay and precession rates for noble-gas spins. 

The responses to oscillation input signals along transverse direction can be obtained by solving the coupled Bloch equations. The response to oscillating magnetic field $B_{y} = B_\text{y0} \cos(\omega t)$ is  
\begin{equation} 
\begin{split}
 P^\text{e}_{{x}( B_{y} )} &= {K_{ B_{y} }} B_\text{y0} \cos (\omega t + {\phi _{ B_{y} }}),\\ 
 K_{ B_{y} } & \approx \frac{{ \gamma_\text{e}  P_{z}^\text{e} }}{Q}\frac{\omega }{{\left\{ {{{\left( { {\omega}^\text{n}_{z} {\omega}^\text{e}_{z}  - {\omega ^2}} \right)}^2} + {{ \left(  {R_2^\text{e} /Q \times \omega}  \right)}^2}} \right\} }^{1/2}},\\
 \label{eq.scalefac_By}
 \end{split}
\end{equation}
where $K_{ B_{y} }$ is the scale factor,  ${\omega}^\text{n}_{z}= \gamma_\text{n} {B}^\text{n}_{z} $, and ${\omega}^\text{e}_{z}= \gamma_\text{e} {B}^\text{e}_{z}/Q$.

The response to oscillation exotic field coupled to noble-gas spins $b_{x}^\text{n} = b_\text{x0}^{\rm n} \cos (\omega t)$ is  
\begin{equation} 
\begin{split}
 P^\text{e}_{{x}( b_{x}^\text{n} )} &= {K_{ b_{x}^\text{n}  }} b_\text{x0}^\text{n} \cos (\omega t + {\phi _{ b_{x}^\text{n}  }}),\\ 
 K_{ b_{x}^\text{n} } &  \approx \frac{{ \gamma_\text{e}  P_{z}^\text{e} }}{Q}\frac{{\omega}^\text{n}_{z} }{{\left\{ {{{\left( {{ {\omega}^\text{n}_{z}} {\omega}^\text{e}_{z}  - {\omega ^2}} \right)}^2} + {{ \left( R_2^\text{e} /Q  \times \omega \right)}^2}} \right\}}^{1/2}}.
 \label{eq.scalefac_bxn}
 \end{split}
\end{equation}

The scale factors for magnetic field $B_{y}$ and exotic field $b_{x}^\text{n}$ have the following relation $K_{ B_{y} }= K_{ b_{x}^\text{n} } \omega / {{\omega}^\text{n}_{z}}$. As a result, the response to oscillating magnetic fields can be used to calibrate the response to exotic fields, which is a conventional calibration method in atomic magnetometer.
\\

\noindent \textbf{Axion Signal Analysis}

\noindent \textbf{\textit{DM Signal Calculations}}

The signal measured in a time sequence can be expressed as the projection of this pseudomagnetic field in the direction of the sensitive axis $\hat{\mathbf{m}}$ at discrete points in time:
    \begin{equation}
    \beta_j = \frac{g_{aNN}}{\gamma_{\rm n}} \, \boldsymbol{\nabla}a (j\Delta t) \cdot \hat{\mathbf{m}}(j \Delta t) \,,
    \label{eq:beta_def}
    \end{equation}
where $\Delta t = 1/(3598\,{\rm Hz})$ is the sampling interval time, and $j$ is an integer index, with $j=N$ corresponding to the total measurement time. As a result of the Earth rotation, the sensitive axis $\hat{m}(t)$ changes with time. The daily modulation effect caused by the Earth rotation can be expressed as
\begin{equation}
\hat{\mathbf{m}}_i(j \Delta t)\approx \mathbf{C}_i \cos(\omega_{\rm E} j \Delta t +\theta_i) + \mathbf{D}_i\,,
\label{eq:axis_approx}
\end{equation}
where $\omega_{\rm E} $ is the angular frequency of the Earth rotation, and the parameters $\mathbf{C}_i$, $\mathbf{D}_i$, and $\theta_i$ are determined by the location, the sensitive axis, and the starting time of the experiment respectively. 
The index $i$ runs from an orthonormal coordinate system $\{  \hat{u}, \hat{v}, \hat{s} \}$ \cite{lee2022laboratory}, where $\hat{v}$ is parallel to the Earth velocity respect to the Sun. The parameters for Dataset 1 and 2 are given in Table~\ref{tab:setups}.

\begin{table}[htb]
\centering
\begin{tabular}{|c|c|c|c|c|c|}
    \hline
    &$\mathbf{C}_i$&$\mathbf{D}_i$  &$\theta_{i}$ (DS-1)  & $\theta_{i}$ (DS-2) & $\theta_{i}$ (DS-3)\\
    \hline
	$\hat u$& 0.89 & 0 & 0.37 & 2.0 & -2.9\\
	\hline
	$\hat v$& 0.67 & 0 & $-0.59$ & 1.0 & 2.5\\
	\hline
	$\hat s$& 0.88 & 0 & $-1.5$ & 0.12 & 1.6\\
	\hline
	\end{tabular}
	\caption{The parameters for the experiment setup. $C_i$ and $D_i $ are the same for Dataset (DS) 1, 2 and 3, because they are taken at nearly the same location. $D_i = 0$ because the sensitive axis points to the West. The phases $\theta_{i}$ (DS-1, DS-2, DS-3) are determined by the starting time of each data set.}
\label{tab:setups}
    \end{table}

  We calibrated the system every four hours to monitor the status during data acquisition. The system was verified to be stable in Ref.~\cite{wei2022constraints}. For higher-frequency signal, we carried out a 209-hr measurement with an $86\%$ duty cycle (Dataset 1), resulting in a measured power spectral density (PSD) data step of $\Delta f = 1/(180\,{\rm hr}) =1.54\, \rm{\mu Hz}$. Additionally, we performed a 4-hr continuous measurement with a nearly identical setup deployed in an underground laboratory (Dataset 2), which provided better suppression of vibrations and power-line interference. Both measurements were conducted at a fixed bias magnetic field in the HSR regime. Subsequently, we transformed the two data sets into the frequency domain using non-uniform Fast Fourier Transformation (FFT). Since the data have gaps, one has to use the non-uniform FFT.
  To accommodate memory constraints (128\,GB of random access memory in the computer we used), we had to down-sample the data by a factor of two, which limited the highest frequency it could accommodate to 900\,Hz. Since there is no down-sampling in Dataset 2, it can cover the frequency range up to $f_s/2$; we studied the frequencies up to  1000\,Hz and Dataset 2 was the only one with which we could access  the [900, 1000]\,Hz range. We employed the log-likelihood ratio (LLR) test to analyze the two data sets separately in the frequency domain for the axion-neutron $g_{ann}$ coupling scenario, accounting for the stochastic effect \cite{lee2022laboratory} 
  The effective coupling between axion and the nucleus $N$ is given by $g_{aNN} = \xi_{\rm n}^{\rm Ne} g_{{ann}} + \xi_{\rm p}^{\rm Ne} g_{ {app}}$, where $\xi_{\rm n}^{\rm Ne} = 0.58/3$ and $\xi_{\rm p}^{\rm Ne} = 0.04/3$ are the fractions of spin-polarization for neutron and proton in ${}^{21}{\rm Ne}$~\cite{brown2017nuclear, Almasi:2018cob}. The factor of 3 is a rescaling factor due to the ${}^{21}$Ne spin $3/2$.

To measure low-frequency signals, where magnetic noise is particularly significant, we use the SC regime \cite{kornack2005nuclear}. In this regime, the noble-gas magnetization automatically compensates low-frequency magnetic fields, leaving alkali spins protected from magnetic noise. The K-$^{3}$He SC comagnetometers have shown ultrahigh sensitivity of about $\rm{1\,fT/Hz^{1/2}}$ \cite{Vasilakis:2008yn}. In our experiment, we use $^{21}$Ne whose gyromagnetic ratio, $\gamma_{\rm Ne} = (2\pi)\times 3.36\,{\rm MHz/T}$, is one order of magnitude smaller than that of $^{3}$He. This results in higher sensitivity to exotic field under the same noise level \cite{Wei:2022ggs}. The SC comagnetometer can be calibrated by oscillation magnetic field \cite{Brown:2010dt} using the residual response of the SC magnetometer proportional to the frequency of the oscillation at low frequencies.  The frequency response to exotic fields coupling to noble-gas nuclear spins is then calibrated by that of magnetic field, in the same way as it is done in the HSR regime. 

The SC regime data were taken over 146\,hr, which is considerably shorter than the 40-day duration of the K-$^{3}$He data taking in Ref. \cite{lee2022laboratory}. 
To make the most of the available data, we calculate the power spectral density (PSD) 
for each hour separately. We then removed the lowest-quality $10\%$ of the data (specifically, one-hour long data segments showing the largest fluctuations), resulting in a final dataset of 132\,hr for the analysis. We note that this procedure suppresses transient stochastic signals that are being searched for in experiments like GNOME \cite{afach2024can}. The analysis of SC data is similar to that for the HSR data and we have derived DM constraints in the frequency range of [0.01, 6]\,Hz.
\\

\noindent \textit{\textbf{Data Preparation} }

The information regarding the start date, end date, and measurement duration for the three datasets is presented in Table~\ref{tab:span}. In Dataset 2 for HSR, we utilized a distinct apparatus situated in an 8-meter-deep underground basement. This facility was originally designed as an air-defense structure and includes a substantial cistern, which contributes to temperature stability and reduced vibration noise compared to Dataset 1. Unfortunately, we were unable to extend the measurement duration of Dataset 2 due to ongoing construction and renovation activities. However, once these construction efforts are completed, we intend to relocate our future experiments to this deep underground site.

\begin{table}[htp]
	\centering
	\begin{tabular}{c|c|c| c}
		\hline
		\hline
		Dataset   & Start Date & End Date & Duration [hr]\\
		\hline
		HSR (Dataset 1) & 2023/01/16  & 2023/01/25 & 209\\
		\hline
		HSR (Dataset 2) &2023/01/18  & 2023/01/18 & 4\\
		\hline
		SC (Dataset 3)&2023/03/31  &2023/04/09  & 146 \\
		\hline
		\hline
	\end{tabular}
	\caption{Information regarding the three datasets. 
	All data were collected in the laboratory located in Hangzhou, China. 
		}
	\label{tab:span}
\end{table}

In our analysis of the axion dark matter signal, we define the analysis frequency band as detailed in Table~\ref{tab:noiseband}. To effectively determine an upper limit for the axion-nucleon coupling strength through the profile log-likelihood analysis, it is crucial that the analysis band spans the complete axion signal linewidth. Furthermore, it is noteworthy that the Earth rotation introduces a $\pm f_E$ modulation to the signal frequency, with $f_{\rm E}\approx 11.6\,\mu\text{Hz}$ representing its rotational frequency. 

For the low-frequency range, mostly corresponding to the SC regime (Dataset 3), $f_a \in [0.01,~6)$ Hz, the Earth rotation effect becomes significant due to the fact that $f_{\rm E}$ can easily surpass the signal line width, approximately on the order of $\Delta f_a \approx 10^{-6} f_a $. Therefore, the selection of our noise frequency band is critical and should encompass the frequency shift induced by the Earth rotation. 
Moreover, to obtain a more precise assessment of the background noise level, we include approximately 200-300 data bins in the log-likelihood analysis. As a result, we have chosen the range $[f_a-20\Delta f_{\rm E},~f_a+20\Delta f_{\rm E} ]$ as our analysis frequency band for Dataset 3. This range remains adequately narrow to uphold the assumption of white noise in the background.

\begin{table}[htp]
	\centering
	\begin{tabular}{c|c}
		\hline
		\hline
		Axion Frequency ($f_a$ [Hz]) & Frequency band used for analysis\\
		\hline
		$[0.01,~6)$   & $[f_a-20\Delta f_{\rm E},~f_a+20\Delta f_{\rm E} ]$ \\
		\hline
		$[6,~1000)$  & $[f_a-\frac{3}{2}\Delta f_a,~f_a+3\Delta f_a ]$ \\
		\hline
		\hline
	\end{tabular}
	\caption{The frequency bands utilized for the log-likelihood profile analysis, with $f_{\rm E}\approx 11. 6 \mu\text{Hz}$ being Earth rotational frequency,  and $\Delta f_a$ representing the signal linewidth. 
	}
	\label{tab:noiseband}
\end{table}

In the context of the high-frequency range, mostly associated with the HSR regime (Dataset 1 and 2), where $f_a$ falls within the interval $[6,~1000)$ Hz, the impact of Earth rotation becomes smaller and $\Delta f_a \geq f_{\rm E}$. 
For instance, when $f_a$ is approximately 6 Hz, the signal linewidth is estimated to be around $\Delta f_a \approx 10.1 ~\mu{\rm Hz}$, which is roughly the frequency of Earth rotation. For the increased axion frequency, the impact of Earth rotation becomes negligible. Therefore, the choice of bandwidth fully includes the modulation of axion signal linewidth. This band choice is intentionally asymmetric, because the total energy of the axion includes its rest mass energy and the kinetic energy.

In this study, we employed the \texttt{MATLAB} built-in non-uniform FFT algorithm, derived from Refs.~\cite{7953011, dutt1993fast}, which is based on the $z$-transform method. The $z$-transform is a generalization of the Fourier transform for discrete series, making it suitable for handling non-uniform time series data. In Ref.~\cite{munteanu2016effect}, the study showed the consistency of the $z$-transform FFT algorithm with the Lomb-Scargle algorithm, which was used in the K-${}^3$He study~\cite{lee2022laboratory}. We also generated non-uniform random test data to compare both algorithms and found negligible differences in the frequency power spectrum density. As a result, we chose to adopt the \texttt{MATLAB} built-in $z$-transform FFT algorithm for its accuracy and convenience. 
\\

\noindent \textit{\textbf{Likelihood Analysis} }

We now briefly introduce the method for setting the limits below, which is based on Ref.\,\cite{lee2022laboratory}. The time sequence of ${ \beta_1, \beta_2, \cdots, \beta_{N}}$ follows a multivariate normal distribution with zero means due to the uniform random phase $\phi_p$ contained in each ${\bf p}$ mode of the axion, in accordance with the central limit theorem. As a result, the upper limits on axion coupling $g_{N}$ must utilize the statistical properties.
				
The Fourier transform of the time series $\beta_j$ produces a complex variable $\tilde{\beta}_k$, which follows a multivariate normal distribution. The frequency series can be redefined in terms of its real and imaginary parts,
\begin{align}
A_k \equiv \frac{2}{N}{\rm Re}[\tilde{\beta}_k] , \quad B_k \equiv -\frac{2}{N}{\rm Im}[\tilde{\beta}_k].
\end{align}
Since $A_k$ and $B_k$ have zero means, their statistical properties are coded in the covariance matrix ${\rm Cov}\left[ A_{k_1}, A_{k_2} \right]$, ${\rm Cov}\left[ B_{k_1}, A_{k_2} \right] $, $ {\rm Cov}\left[ A_{k_1}, B_{k_2} \right] $ and ${\rm Cov}\left[ B_{k_1}, B_{k_2} \right] $ with $(A/B)_{k_1/k_2} $ being the specific choices of $(A/B)_{k}$. Their calculations are detailed in Ref.~\cite{lee2022laboratory}. Furthermore, if the duration of data taking $T$ is much greater than the coherence time $\tau_a$, the covariance matrix calculation can be significantly simplified.

The experimental background within a sufficiently narrow bandwidth can be modeled as Gaussian white noise with a zero mean and variance $\sigma_b^2$. Let the measured data in the frequency domain be the vector ${\bf d} = \left\{ A_k, B_k\right\} $, with a variance matrix given by $\Sigma = \Sigma_a + \sigma_b^2 \mathbbm{1}$. Here, $\Sigma_a(g_{aNN}, m_a)$ is the covariance matrix of the axion signal, which depends on the coupling $g_{aNN}$ and the axion mass $m_a$, while $\mathbbm{1}$ is the identity matrix corresponding to the Gaussian white noise~\cite{lee2022laboratory}.  
The covariance matrix of the axion signal is explicitly given by  
\begin{align}
    \Sigma_a(g_{aNN}, m_a)_{k,r} = \begin{pmatrix}
{\rm Cov}(A_k, A_r) & {\rm Cov}(A_k, B_r) \\
{\rm Cov}(B_k, A_r) & {\rm Cov}(B_k, B_r)
\end{pmatrix} 
\,,
\end{align}
where the dependence on the frequency bin indices $k$ and $r$ ensures that the full shape of the axion signal is encoded within it. Consequently, the axion signal, including its stochastic effects, is fully captured by $\Sigma_a$. 
Since both the signal and the background are multivariate Gaussian random variables, one can finally construct the likelihood function as \cite{derevianko2018detecting, lee2022laboratory},
\begin{align}
L(\textbf{d}|g_{aNN}, \sigma_b^2)=\frac{1}{\sqrt{(2\pi)^{2N}\mathrm{det}(\Sigma)}} \mathrm{exp}\left(-\frac{1}{2}\textbf{d}^T\Sigma^{-1}\textbf{d}\right).
\end{align}

\noindent \textit{\textbf{Calculation of Axion Limits} }

To set a quantitative limit, one can use the log-likelihood ratio (LLR) below~\cite{lee2022laboratory}
\begin{align}
					{\rm LLR}(g_{aNN}) =
					\begin{cases}
						-2\log\left[\frac{L(g_{aNN},{\tilde{\sigma}_b})}{L(\hat{g}, \hat{\sigma}_b)}\right],&\hat{g}\leq  g_{\rm aNN} \\
						0, &\hat{g}> g_{\rm aNN}
					\end{cases}
\end{align}	where $\tilde{\sigma}_b$ maximizes $L$ for a fixed nucleus coupling $g_{aNN}$. In order to establish an upper limit, ${\rm LLR}=0$ if $\hat{g} > g_{aNN}$. The $95\%$ CL upper limits are set by finding the value of $g_{aNN}$ where ${\rm LLR}(g_{aNN})=2.71$. Finally, we calculate the two sets of data separately and choose the stronger limits on $g_{aNN}$ among the two. By applying the spin polarization fractions in ${}^{21}{\rm Ne}$, the limits on the nucleus coupling $g_{aNN}$ can be converted to nucleon couplings $g_{ann}$ and $g_{app}$, resulting in the final outcome displayed in Fig.\,4 of the main text. To set a continuous limit at axion mass parameter spaces, the limit frequencies are separated by $\Delta f =1/(2\tau_a)$, leading to over 10 millions of limits points.
				
For $f_a \in [3,30]$ Hz, the sensitivity to $g_{ann}$ remains approximately constant at a level of about $3 \times 10^{-9}\,{\rm GeV}^{-1}$, while for $f_a \in [30,1000]$ Hz, it is roughly proportional to $f_a^2$. The shape of the curve is determined by the sensitivity to the magnetic field $B_y$ and the ratio between the two responses $K_{B_y}/K_{b_x^{\rm n}} \propto \omega_a$. Neglecting the stochastic effect, the sensitivity to the axion signal at frequency $f_a$ can be estimated from the square root of the data power spectrum, denoted as $\sigma_b$ which is the standard deviation of $B_y$. The $95\%$ confidence level (CL) limit can be obtained by roughly letting $g_{ann} \approx  (2.7 \sigma_B K_{B_y}/K_{b_x^{\rm n}})/ (\xi_{\rm n} \sqrt{\rho_{\rm DM}} v_{\rm DM}/\gamma_{\rm n})$. Therefore, the coupling $g_{ann}$ is proportional to $f_a \sigma_b$. 
In the PSD data, we observed that $\sigma_B$ is proportional to $f_a^{-1}$ for $f_a \in [3,30]$ Hz, flat for $f_a \in [30,50]$ Hz, and proportional to $f_a$ for $f_a \in [50,1000]$ Hz. Including the stochastic effect will enlarge the coupling $g_{ann}$ by a factor of a few, but it has no significant dependence on $f_a$. Therefore, the slope of $g_{ann}$ in the final results matches the simple estimation above quite well.
				
In alkali-noble gas hybrid sensors, the energy resolution for measuring the shifts of the levels of the noble-gas nuclear spins is typically much higher than that for alkali spins, primarily because the higher atom density, longer coherence time and smaller magnetic moment \cite{Terrano:2021zyh}. Old K-$^3$He comagnetometer data can set limits on the axion neutron coupling in the frequency below 10\,Hz. There are two results interpreted by the same experiment. In Fig.\,4 of the main text, we only plot the limits using the original data and properly treated the stochastic effects of axion\,\cite{lee2022laboratory}, and ignore the one that indirectly interpreted the results from published power spectra\,\cite{Bloch:2019lcy}. Our SC data provides comparable and even stronger limits to the K-$^3{\rm He}$ study~\cite{lee2022laboratory} for the axion-neutron coupling $g_{ann}$.
In terms of the axion-proton coupling $g_{app}$, we have achieved the most stringent terrestrial constraints across an extensive frequency range of [$10^{-2}$, $7\times 10^2$]\,Hz. The K-$^3{\rm He}$ results can also provide limits on the axion-proton coupling. However, considering the spin polarization fractions in $^3{\rm He}$, with $\xi_{\rm n}^{\rm{^3}He}=87\%$ and $\xi_{\rm p}^{\rm{^3}He}=-2.7\%$~\cite{Vasilakis:2008yn}, we found that their constraints are approximately two times weaker than ours.

\begin{figure}[htb]
    \centering
    \includegraphics[width= 0.98 \linewidth]{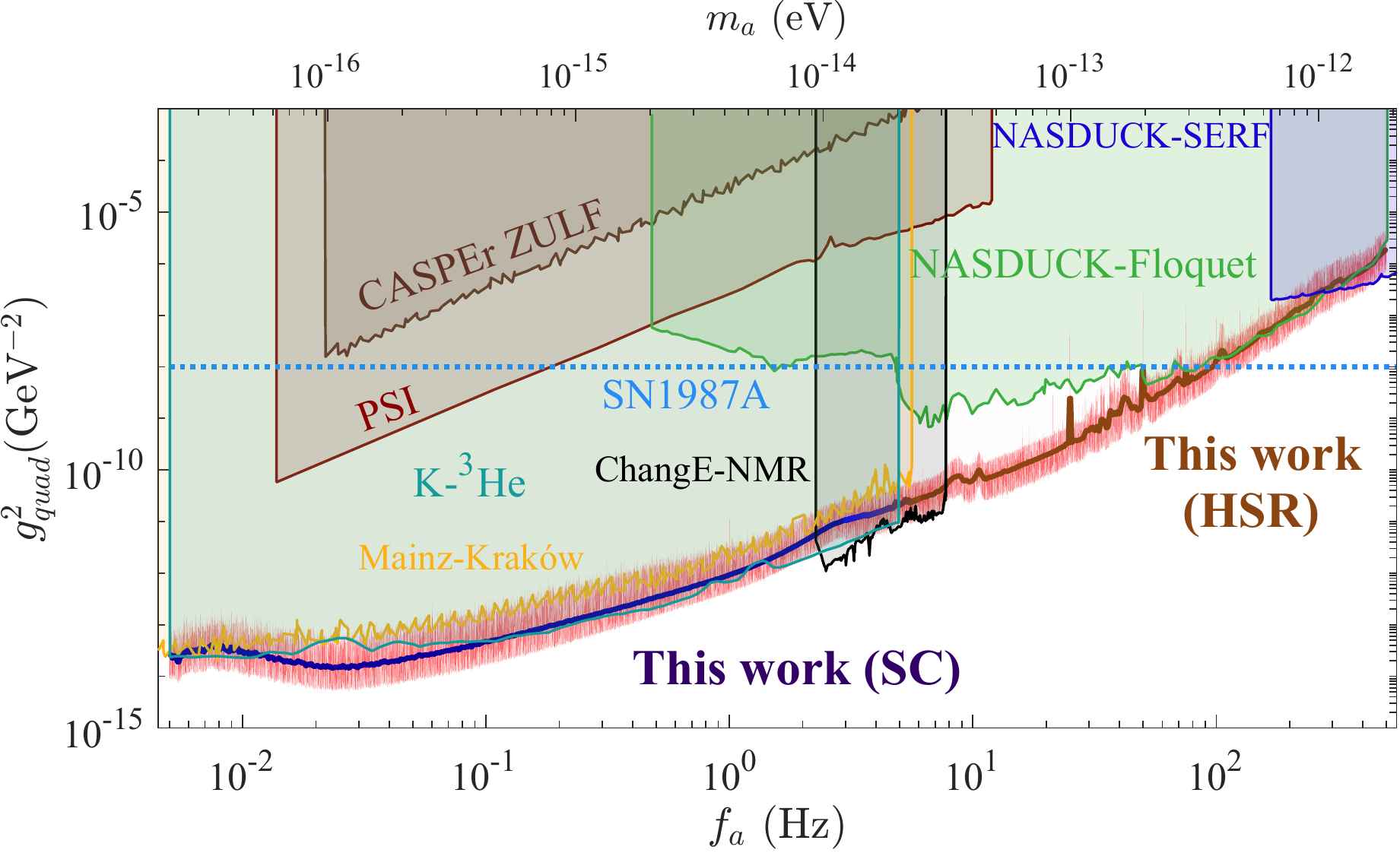}
    \caption{The 95\% C.L. upper limits (depicted by red lines) for the quadratic neutron coupling $g_{\rm quad}^2$ are derived from measurements in both the SC and HSR regimes, using the same labels as in Fig.\,4 of the main text. }
    \label{fig:axion-limits-quad} 
\end{figure}

In addition to the linear axion derivative coupling, we investigated the limits on the quadratic coupling scenario~\cite{Olive:2007aj, Pospelov:2012mt, wu2019search, jiang2021search} following a similar procedure as for linear coupling. The corresponding Hamiltonian can be expressed as:
    \begin{align}
\mathcal{H}_{\rm quad} =  g_{\rm quad}^2 \mathbf{\nabla} a^2\cdot \bf{I}.
    \end{align}
Assuming that the quadratic coupling dominates, we can use the same analysis as above to recast it as results on $g_{\rm quad}$. These are presented in Fig.\,\ref{fig:axion-limits-quad}. As a result of the quadratic coupling, the range of axion frequencies is reduced by a factor of two compared to the linear coupling, resulting in a range of $[5\times 10^{-3}, \, 5\times 10^{2}]$ Hz. The limits surpass the astrophysical constraints by several orders of magnitude for $f_a \lesssim 150$\, Hz. These results also surpass the earlier terrestrial experiments.
\\

\begin{table*}[htb]
	\centering
		\begin{tabular}{cclcc|cclcc}
			\hline
			Frequency [Hz]   &Full-data LLR & Joint-LLR & Sig-Stability & Sig-Shape  &  Frequency [Hz]  &Full-data LLR &  Joint-LLR &   Sig-Stability & Sig-Shape  \\
			\hline
			46.704996 &69.77 & \usym{2713} (68.) & \usym{2713} & \usym{2717} & 81.504437 &  56.57 & \usym{2717} (5.87) & &  \\
			48.391586 &53.88 & \usym{2717} (2.49)& & & 81.722087 & 109.95  & \usym{2717} (14.8) & & \\
			67.129321 &69.04 & \usym{2717} (0.06)& & & 81.754539 & 53.81  & \usym{2717} (1.24)& & \\
			67.143839 &75.89 & \usym{2713} (31.36)& \usym{2717} & & 81.841119  & 60.90 & \usym{2717} (11.3) & & \\
			67.162956 &54.79 & \usym{2717} (18.79)& & & 97.660449 & 79.31  & \usym{2713} (84.27) & \usym{2713} &\usym{2717} \\
			67.164261 &65.76 & \usym{2717} (2.67)& & & 134.141313 & 57.61 & \usym{2717} (6.94)& & \\
			67.239592 &80.46 & \usym{2717} (0)& & & 134.248211 & 66.08 & \usym{2717} (9.1)& & \\
			67.256066 &68.44 & \usym{2717} (9.16)& & & 134.451359 &76.66  & \usym{2717} (7.91)& & \\
			67.299372 &59.97 & \usym{2717} (0)& & & 134.454312 & 64.66 & \usym{2717} (2.27)& & \\
			67.305285 &52.94 & \usym{2717} (0.03)& & & 134.459877 &54.65 & \usym{2717} (9.20)& & \\
			67.372806 &75.25 & \usym{2717} (0.95)& & & 137.940287 & 71.64  & \usym{2717} (0.13)& & \\
			67.378668 &96.18 & \usym{2717} (7.01)& & & 138.303853 & 54.19  & \usym{2717} (4.01)& & \\
			67.384759 &58.44 & \usym{2717} (0)& & & 140.709636 & 70.36 & \usym{2717} (8.71)& & \\
			67.399959 &55.77 & \usym{2717} (0.68)& & & 140.717362 & 59.57 & \usym{2717} (1.47)& & \\
			67.401895 &62.05 & \usym{2717} (0.07)& & &  140.733529 & 58.62 & \usym{2713} (31.01)& \usym{2717} & \\
			67.424104 &55.74 & \usym{2717} (0.11)& & & 140.736620 & 53.61 & \usym{2717} (21.57)& & \\
			67.439142 &68.60 & \usym{2717} (1.72)& & & 140.791675 & 72.28  & \usym{2717} (3.85)& & \\
			67.447232 &70.18 & \usym{2717} (9.51)& & & 140.794648 & 52.74 & \usym{2717} (17.40)& & \\
			67.457887 &90.58 & \usym{2717} (6.36)& & & 141.246730 & 62.39 & \usym{2717} (16.18)& & \\
			72.782530 &60.67 & \usym{2717} (12.03)& & & 141.701699& 96.47  & \usym{2713} (23.71)& \usym{2717} & \\
			72.785297 &52.33 & \usym{2717} (16.78)& & & 141.952935 & 63.83 & \usym{2717} (0.05)& & \\
			72.788802 &56.44 & \usym{2717} (1.17)& & & 142.624594 & 79.79 & \usym{2717} (6.57)& & \\
			72.797656 &53.53 & \usym{2717} (3.07)& & & 142.729812 & 65.91 & \usym{2713} (27.34)&\usym{2717} & \\
			72.803745 &61.79 & \usym{2717} (0.3)& & & 143.372442 & 62.12 & \usym{2717} (9.34)& & \\
			72.823120 &58.99 & \usym{2717} (0)& & & 143.969920 & 55.22 & \usym{2717} (2.41)& & \\
			72.825273 &52.13 & \usym{2717} (10.92)& & & 157.812307& 63.13  & \usym{2717} (4.07)& & \\
			72.839546 &57.13 & \usym{2717} (10.5)& & & 159.152266 & 68.58 & \usym{2713} (48.39)& \usym{2717}& \\
			72.869210 &83.03 & \usym{2717} (0.53)& & & 161.744217 & 64.57 & \usym{2713} (76.18)& \usym{2713} & \usym{2717} \\
			72.871118 &52.66 & \usym{2717} (6.26)& & & 186.715234 & 75.55 & \usym{2713} (43.96)& \usym{2717} & \\
			72.874196 &80.23 & \usym{2717} (3.91)& & & 188.360579 & 86.56 & \usym{2713} (47.25)& \usym{2717}& \\
			72.921797 &60.32 & \usym{2717} (0)& & & 363.317150 & 120.56 & \usym{2713} (142.83)& \usym{2713} & \usym{2717} \\
			\hline
		\end{tabular}
    \caption{The flowchart illustrates the discrimination process for the 62 candidates, involving the two-stage joint-LLR test (threshold at 23.7), signal-strength-stability test, and signal-shape test. 
    In the second column, the full-data LLR results are presented for comparison with the joint-LLR results. None of the candidates passed all three tests. 
	}
	\label{tab:flow_chart-final-1}
\end{table*}

\noindent \textbf{Post-Analysis of the Possible Candidates.} 

We explore if there are possible axion DM signal in the data. For a quantitative analysis of possible axion candidates, we define the following test statistics to test the significance of a best-fit signal compared to the background only model \cite{lee2022laboratory},
\begin{align}
	{\rm LLR}_\text{discover} =
	\begin{cases}
		-2\ln\left[\frac{L(0,{\tilde{\sigma}_b})}{L(\hat{g}, \hat{\sigma}_b)}\right],&\hat{g}\geq 0 \\
		0, &\hat{g} < 0,
	\end{cases}
	\label{eq:5-sigma-discovery-v1}
\end{align}
where $\tilde{\sigma}_b$ maximizes ${\rm LLR}$ without the signal, while $\hat{g}$ and $\hat{\sigma}_b$ maximize ${\rm LLR}$ for a two variable marginalization.

Taking into account the look-elsewhere effect, we conservatively estimate that ${\rm LLR} > 52.1$ corresponds to a one-sided global significance of $5\sigma$. We found that out of the $6.8$ million tested masses, 36600 candidates exceeded the $5\sigma$ confidence level, which accounts for approximately $0.5\%$ of Dataset 1. It is evident that there is systematic noise present in Dataset 1, which is not adequately accounted for by the assumption of random Gaussian noise. 
Regarding the potential sources of systematic noise in Dataset 1, significant power grid noise in the harmonics of 50 Hz was observed when examining the power spectrum density of the data. Furthermore, due to the extended measurement time, the drift of the power line noise contributes to broadening the noise range. Additionally, at low frequencies, vibration peaks were detected around 6 Hz and from 17 to 23 Hz, as confirmed using a commercial seismometer.

Fortunately, Dataset 2 possesses superior noise control within the clean underground environment. Initially, we cross-check the candidate frequencies for their $5\sigma$ significance level in Dataset 2, incorporating the look-elsewhere effect. No candidates are identified beyond the $5\sigma$ level, indicating the absence of significant systematic noise in Dataset 2. In determining the final sensitivity result, the combined sensitivity is derived by selecting the superior sensitivity between Dataset 1 and Dataset 2.

Among the 36600 candidates, 62 exhibit better sensitivity in Dataset 1 than in Dataset 2 ($g_{\rm aNN}^{\rm D1} < g_{\rm aNN}^{\rm D2}$). Consequently, the remaining candidates are replaced by the improved limits from Dataset 2, aligning with the random Gaussian noise assumption and ensuring the robustness of their $95\%$ exclusion limits. This process leads to a significant reduction in the number of high LLR points, now totaling only 62, constituting a small fraction of approximately $10^{-5}$.

Subsequently, the remaining 62 candidates undergo further scrutiny using Dataset 1. Three distinct tests are employed to explore their potential as dark matter candidates: the two-stage joint-LLR test, the signal-strength-stability test, and the signal-shape analysis.
\\

\begin{table*}[htb]
	\centering
		\begin{tabular}{c|c|c|c}
			\hline
			\hline
			Frequency [Hz]     & Dataset 1A $[{\rm GeV}^{-1}$]& Dataset 1B  [${\rm GeV}^{-1}$] & Signal Stability? \\
			\hline
			46.704996  & $[2.74\times 10^{-9},~4.26\times 10^{-9}]$ & $[2.52\times 10^{-9},~4.18\times 10^{-9}]$  &\usym{2713} \\
			\hline
			67.143839   & $[2.17\times 10^{-9},~3.97\times 10^{-9}]$ & $[7.66\times 10^{-9},~1.10\times 10^{-8}]$  &\usym{2717} \\
			\hline
			97.660449   & $[6.79\times 10^{-9},1.02~\times 10^{-8}]$ & $[6.93\times 10^{-9},~1.02\times 10^{-8}]$  &\usym{2713} \\
			\hline
			140.733529   & $[1.37\times 10^{-8},~1.89\times 10^{-8}]$ & $[{\rm N/A},~7.59\times 10^{-9}]$  &\usym{2717} \\
			\hline
			141.701699    & $[1.85\times 10^{-8},~2.49\times 10^{-8}]$ & $[6.25\times 10^{-10},~8.16\times 10^{-9}]$  &\usym{2717} \\
			\hline
			142.729812   & $[1.95\times 10^{-8},~2.83\times 10^{-8}]$ & $[5.66\times 10^{-9},~1.07\times 10^{-8}]$  &\usym{2717} \\
			\hline
			159.152266    & $[2.03\times 10^{-8},~2.91\times 10^{-8}]$ & $[9.46\times 10^{-9},~1.62\times 10^{-8}]$  &\usym{2717} \\
			\hline
			161.744217    & $[1.32\times 10^{-8},~2.02\times 10^{-8}]$ & $[1.35\times 10^{-8},~1.99\times 10^{-8}]$  &\usym{2713} \\
			\hline
			186.715234   & $[{\rm N/A },~1.27\times 10^{-8}]$ & $[2.51\times 10^{-8},~3.29\times 10^{-8}]$  &\usym{2717} \\
			\hline
			188.360579   & $[{\rm N/A },~1.57\times 10^{-8}]$ & $[3.1\times 10^{-8},~4.04\times 10^{-8}]$  &\usym{2717} \\
			\hline
			363.317150   & $[8.20\times 10^{-8},~1.13\times 10^{-7}]$ & $[9.34\times 10^{-8},~1.21\times 10^{-7}]$  &\usym{2713} \\
			\hline\hline
	\end{tabular}
	\caption{The table presents the signal-strength intervals calculated at the $90\%$ confidence level for the 11 remaining candidates after the joint-LLR two-stage test. As recommended in Ref.~\cite{lee2022laboratory}, if the signal strength intervals for Dataset 1A and 1B do not overlap, the possibility of the candidate being a dark matter signal can be excluded. The notation ``N/A'' means that, no solution is found for the lower threshold $g_{\rm aNN}^{\rm lower}$ at $90\%$ confidence level.
	}
	\label{tab:CLintervals}
\end{table*}

(1) \textbf{The Two-Stage Joint-LLR test}: In this test, the objective is to ascertain whether the identified candidates can be attributed to transient systematic noise. Therefore, we divided Dataset 1 into two parts, each with an equal duration of 90 hours. For two datasets, 1A and 1B, of the same size, these two datasets can be considered independent, given that the time span of 90 hours significantly exceeds the coherence time of the smallest axion frequency among the 62 candidates. Thus, the joint-likelihood function can be constructed as follows:
\begin{align}
	L(g_{\rm aNN},\sigma_{b1A},\sigma_{b1B})\equiv L_{1A}(g_{\rm aNN},\sigma_{b1A})\times L_{1B}(g_{\rm aNN},\sigma_{b1B})\,,
\end{align}
where $\sigma_{b1A}$ and $\sigma_{b1B}$ represent the noise levels of the two datasets. Following a similar approach as in the main text, the log-likelihood ratio for joint-significance is defined as:
\begin{align}
	{\rm LLR}_\text{discover}^{\rm joint} =
	\begin{cases}
		-2\ln\left[\frac{L(0,\ \tilde{\sigma}_{b1A},\ \tilde{\sigma}_{b1B})}{L(\hat{g},\ \hat{\sigma}_{b1A},\ \hat{\sigma}_{b1B})}\right],&\hat{g}\geq 0, \\
		0, &\hat{g} < 0,
	\end{cases}
\end{align}
where the hats and the tildes represent the marginalization of these parameters in the denominator and numerator, respectively.

The joint-LLR, as defined, serves the purpose of distinguishing transient noise peaks from persistent signal-like peaks using the two-stage datasets. In the flowchart Table~\ref{tab:flow_chart-final-1}, we have tabulated the joint-LLR results alongside the full-data LLR for clear comparison. It is evident that many of the joint-LLR values are significantly lower when compared to the LLR, suggesting that these candidates likely originate from transient noise backgrounds.

It is worth noting that at specific test frequencies, we have obtained nearly zero joint-LLR results. These candidates exhibit a common characteristic: the noise levels of the two datasets vary significantly, and the noise peak at that particular test frequency only appears in the noisier dataset. Consequently, the joint-LLR is calculated to be very small, approaching zero, as no signal is required to fit the data under these circumstances. 

In Table~\ref{tab:flow_chart-final-1}, 11 candidates out of 62 have successfully passed the joint-LLR test, with joint-LLR values exceeding the criterion of 23.7.  This threshold is notably smaller than the ${\rm LLR}_{\rm discover}>52.1$ required in the full data LLR analysis. This is because, with the identification of the frequencies of the 62 candidates, the look-elsewhere effect no longer needs to be considered in the same dataset. 
\\

\begin{figure}[htp!]
	\centering
	\includegraphics[width= 0.99  \linewidth]{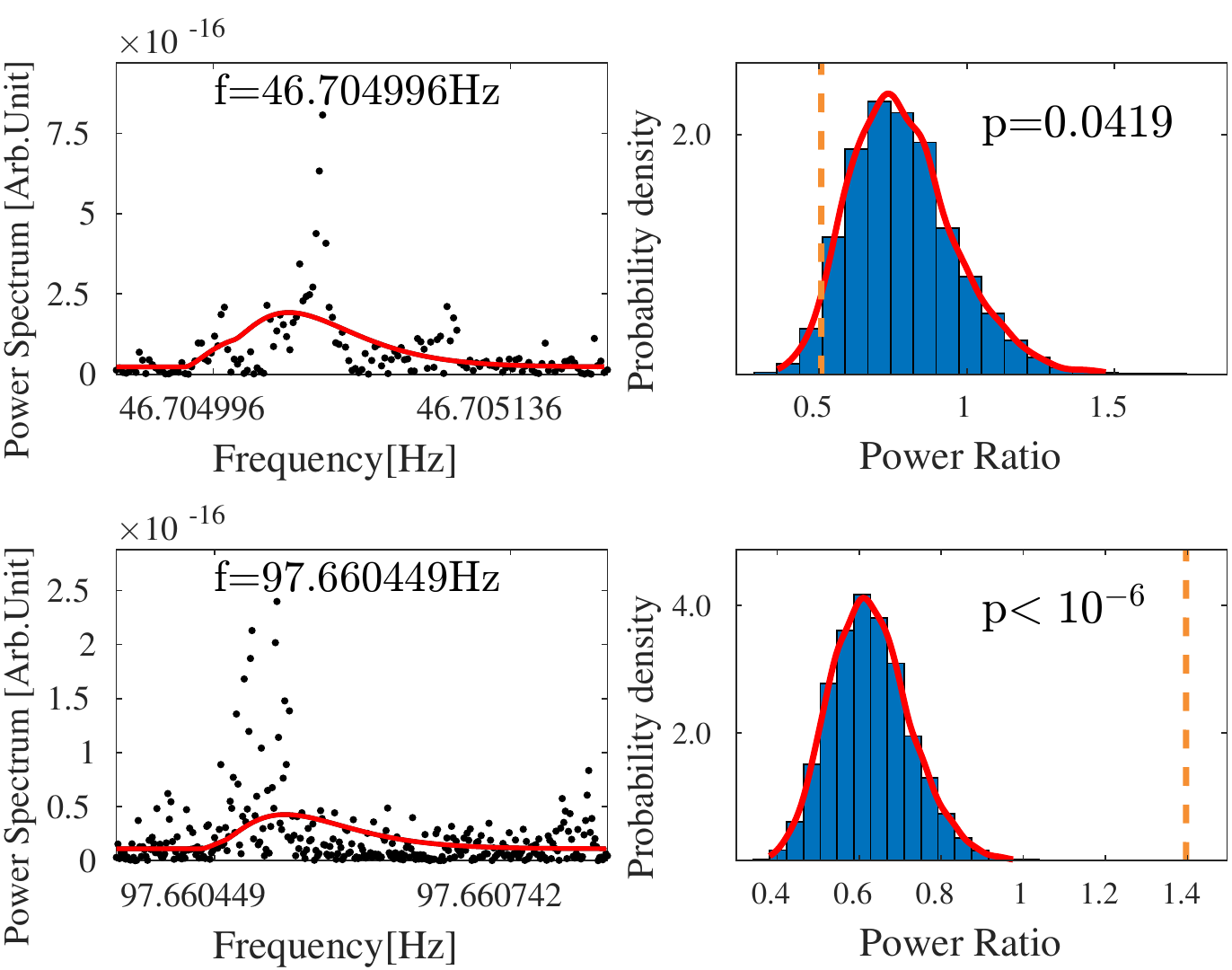}
	\includegraphics[width= 0.99 \linewidth]{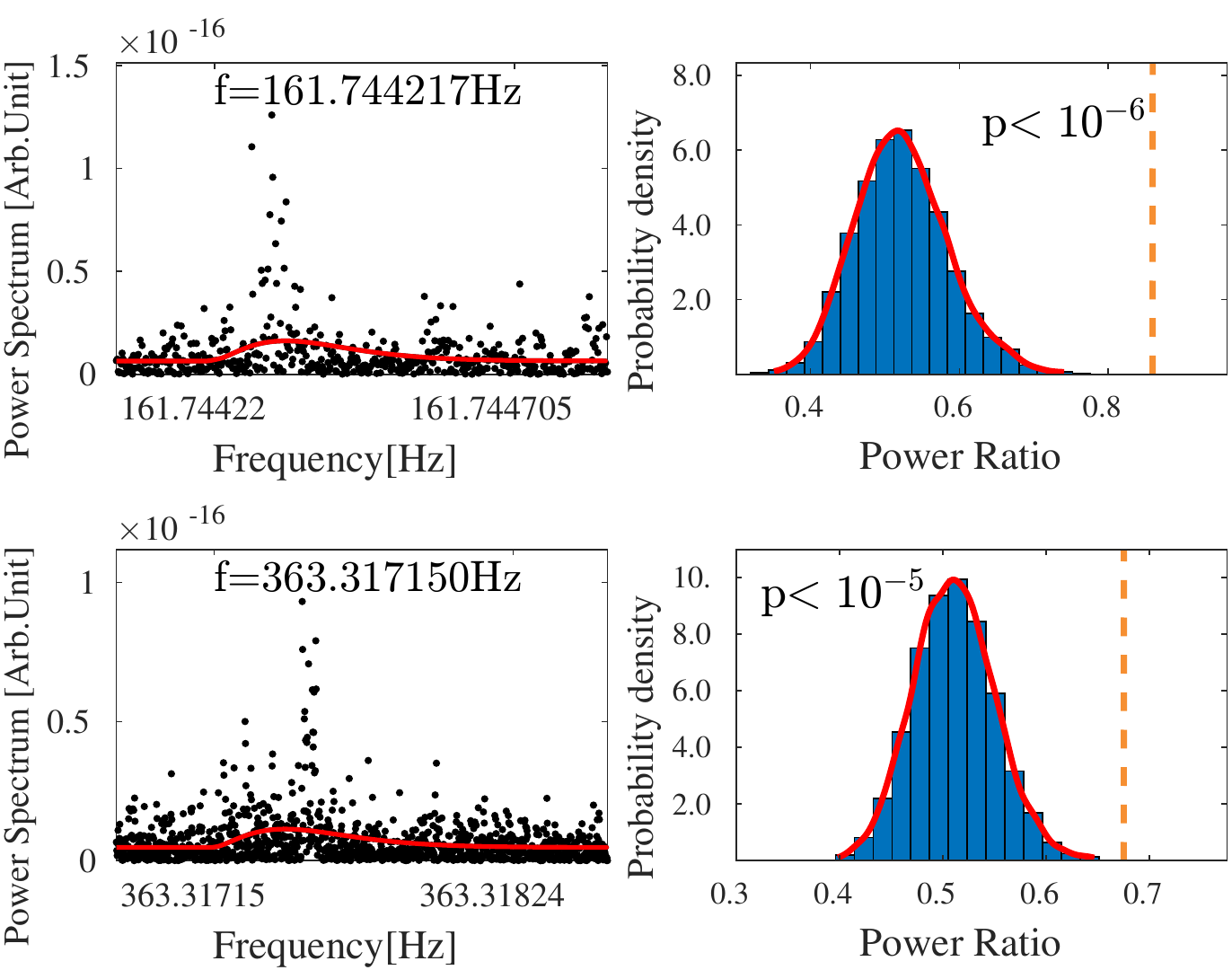} 
	\caption{
	\textit{Left panel}: The black dots represent the power spectra of the data. The expected power spectra of the axion signals are shown as red solid lines. 
 	\textit{Right panel}: The distribution of the power ratio for the axion candidate is determined through Monte-Carlo simulations and is represented as a histogram. The orange lines in the figure indicate the power ratio observed in the data, and the $p$-value quantifies the likelihood of obtaining this power ratio based on the distribution. The four candidates thus fail the $95\%$ confidence-level test (the one around 46.7\,Hz--only marginally) and are attributed to monochromatic noise which does not heve the expected signal shape. As a result, these candidates are rejected as potential axion candidates at the $95\%$ confidence level. }
	\label{fig:lineshapetest} 
\end{figure}

(2) \textbf{The Signal-Strength-Stability Test}:
In accordance with the analysis of Ref.\,\cite{lee2022laboratory}, another method to distinguish noise from signal involves verifying whether the best-fit signal strength, denoted as $\hat{g}_{\rm aNN}$ for Dataset 1A and 1B, remains consistent throughout the entire measurement period. In this context, the best-fit signal strengths must overlap with each other. To derive the confidence interval for the best-fit $\hat{g}_{\rm aNN}$, we introduce a slightly different test statistic in comparison to Eq.\,\eqref{eq:5-sigma-discovery-v1}:
\begin{align}
	{\rm LLR}_{\rm CL}(g_{\rm aNN}) =
	-2\log\left[\frac{L(g_{\rm aNN},{\tilde{\sigma}_b})}{L(\hat{g}_{\rm aNN}, \hat{\sigma}_b)}\right]\,.
	\label{eq:teststatisticCL}
\end{align}

It is important to note that this LLR is not set to zero for $\hat{g}_{\rm aNN}>g_{\rm aNN}$. This particular definition of LLR can reveal any inconsistency between the hypothesized signal strength $g_{\rm aNN}$ and the best-fit $\hat{g}_{\rm aNN}$, regardless of whether $\hat{g}_{\rm aNN}>g_{\rm aNN}$ or $\hat{g}_{\rm aNN}<g_{\rm aNN}$. For the true signal value $g_{\rm aNN}$, the cumulative distribution function (CDF) of ${\rm LLR}_{\rm CL}(g_{\rm aNN})$ is described by the error function~\cite{Cowan:2010js}:
\begin{align}
	P({\rm LLR}_{\rm CL}(g_{\rm aNN})\leq y)={\rm erf}\left (\sqrt{\frac{y}{2}}\right )\,. 
\end{align}

Hence, the upper and lower bounds of the confidence intervals for a confidence level ${\rm CL}_0$ can be determined by the equations: $P[{\rm LLR}_{\rm CL}(g_{\rm aNN})\geq {\rm LLR}_{\rm CL}(g_{\rm aNN}^{\rm upper/lower})]=(1-{\rm CL_0})/2$ for $g_{\rm aNN}^{\rm lower}<\hat{g}_{\rm aNN}$ and $g_{\rm aNN}^{\rm upper}>\hat{g}_{\rm aNN}$, respectively. The results for $90\%$ confidence level intervals for the 11 test frequencies are presented in Table~\ref{tab:CLintervals}. Test frequencies with non-overlapping confidence level intervals can be excluded as potential axion signals at a $90\%$ confidence level. Our analysis shows that there are four frequencies, namely $f_a=46.704996,~97.660449,~ 161.744217,~ 363.317150$\,Hz, that have passed the signal-strength-stability test.
\\

(3) \textbf{The Signal-Shape Analysis}:
In the final step, we need to employ signal shape information to distinguish between a persistent systematic noise peak and an axion dark matter signal. This discrimination can be achieved by either incorporating a monochromatic noise peak into the background model to assess the statistical possibility, as demonstrated in Ref.\,\cite{bloch2022new}, because in the analysis the background was assumed to be Gaussian white noise. Alternatively, one can analyze the shape of the power spectrum density of the data and compare it with the axion signal lineshape using Monte-Carlo methods, as shown in Ref.\,\cite{lee2022laboratory}.

Following the latter approach presented in Ref.\,\cite{lee2022laboratory}, a power ratio $PR$ can be defined as $PR \equiv \text{P}_1/\text{P}_2$, where $\text{P}_1$ represents the  power in the frequency band $[f_a,~f_a+\Delta f_a/2]$. 
In contrast, $\text{P}_2$ is the power in the subsequent frequency band $[f_a+\Delta f_a/2,~f_a+2\Delta f_a]$, which characterizes the power in the high-energy tail. This power ratio $PR$ reflects the lineshape of the axion dark matter signal.

In the left panel of Fig.\,\ref{fig:lineshapetest}, the black dots represent the power spectra of the data while the expected power spectra of the axion signals are shown as red solid lines. We have conducted 10000 Monte-Carlo simulations for each of the four candidate frequencies, considering them as axion-dark-matter candidates. In the right panel of  Fig.\,\ref{fig:lineshapetest}, we present the simulated distribution of the power ratio $PR$, depicted as blue histograms. Using the real data from the left panel, one can calculate the true power ratio of the data, shown as the orange vertical line. The $p$-value quantifies the likelihood of obtaining this power ratio based on the distribution.
If the $p$-value is smaller than $0.05$, we conclude that the four test frequencies are unlikely to be axion candidates at a $95\%$ confidence level. 

Three of the candidates exhibited extremely small $p$-values, less than $10^{-5}$, indicating they are inconsistent with the axion dark matter signal shape and are most likely to be persistent systematic noise peaks. The candidate at a frequency of $f_a = 46.704996$ Hz had a $p$-value of 0.0419, still excluded at the $95\%$ confidence level though marginally.

In summary, after the two-stage joint-LLR test, the signal-strength-stability test, and the signal-shape analysis, there are no candidates left as potential axion dark matter  candidates. However, it is worth noting that the candidate at a frequency of $f_a = 46.704996$ Hz is marginally excluded, and further tests are encouraged.

\section*{Acknowledgement}

We would like to thank Junyi Lee, Yevgeny Kats and Itay Bloch for helpful discussions. 
The work of KW, WQ and JCF is supported by National Science Foundation of China (NSFC) under Grants No. 62203030 and 61925301 for Distinguished Young Scholars, also supported by NSFC-CAS under Grant No. XK2023XXC002, and by the Innovation Program for Quantum Science and Technology under Grant 2021ZD0300401. The work of JL is supported by NSFC under Grant No. 12475103, No. 12235001, No. 12075005, and by Peking University under startup Grant No. 7101502458. The work of XPW is supported by the NSFC under Grant No.12375095  and the Fundamental Research Funds for the Central Universities. The work of WJ and DB is supported by the DFG Project ID 390831469: EXC 2118 (PRISMA+ Cluster of Excellence), by the German Federal Ministry of Education and Research
(BMBF) within the Quantumtechnologien program
(Grant No. 13N15064), by the COST Action within the project COSMIC WISPers (Grant No. CA21106), and by the QuantERA project
LEMAQUME (DFG Project No. 500314265).

\bibliographystyle{utphys}
\bibliography{refs}

\end{document}